\documentclass[12pt]{article}

\catcode`\@=11
\def\marginnote#1{}
\newcount\hour
\newcount\minute
\newtoks\amorpm
\hour=\time\divide\hour by60
\minute=\time{\multiply\hour by60 \global\advance\minute by-\hour}
\edef\standardtime{{\ifnum\hour<12 \global\amorpm={am}%
        \else\global\amorpm={pm}\advance\hour by-12 \fi
        \ifnum\hour=0 \hour=12 \fi
        \number\hour:\ifnum\minute<10 0\fi\number\minute\the\amorpm}}
\edef\militarytime{\number\hour:\ifnum\minute<10 0\fi\number\minute}
\def\draftlabel#1{{\@bsphack\if@filesw {\let\thepage\relax
   \xdef\@gtempa{\write\@auxout{\string
      \newlabel{#1}{{\@currentlabel}{\thepage}}}}}\@gtempa
   \if@nobreak \ifvmode\nobreak\fi\fi\fi\@esphack}
        \gdef\@eqnlabel{#1}}
\def\@eqnlabel{}
\def\@vacuum{}
\def\draftmarginnote#1{\marginpar{\raggedright\scriptsize\tt#1}}
\def\draft{\oddsidemargin -.5truein
        \def\@oddfoot{\sl preliminary draft \hfil
        \rm\thepage\hfil\sl\today\quad\mil922 itarytime}
        \let\@evenfoot\@oddfoot \overfullrule 3pt
        \let\label=\draftlabel
        \let\marginnote=\draftmarginnote
   \def\@eqnnum{(\theequation)\rlap{\kern\marginparsep\tt\@eqnlabel}%
\global\let\@eqnlabel\@vacuum}  }

\def\preprint{\twocolumn\sloppy\flushbottom\parindent 1em
        \leftmargini 2em\leftmarginv .5em\leftmarginvi .5em
        \oddsidemargin -.5in    \evensidemargin -.5in
        \columnsep 15mm \footheight 0pt
        \textwidth 250mmin      \topmargin  -.4in
        \headheight 12pt \topskip .4in
        \textheight 175mm
        \footskip 0pt
        \def\@oddhead{\thepage\hfil\addtocounter{page}{1}\thepage}
        \let\@evenhead\@oddhead \def\@oddfoot{} \def\@evenfoot{} }

\def\titlepage{\@restonecolfalse\if@twocolumn\@restonecoltrue\onecolumn
     \else \newpage \fi \thispagestyle{empty}\c@page\z@
        \def\thefootnote{\fnsymbol{footnote}} }

\def\endtitlepage{\if@restonecol\twocolumn \else  \fi
        \def\thefootnote{\arabic{footnote}} \setcounter{footnote}{0}}
\catcode`@=12
\relax
\def\bea{\begin{array}}
\def\bem{\begin{displaymath}}
\def\beq{\begin{equation}}

\def\eea{\end{array}}
\def\eem{\end{displaymath}}
\def\eeq{\end{equation}}
\def\s2w{\sin^2 \theta_W}
\def\Tr{\mathop{\rm Tr}}

\relax
\def\be{\begin{equation}}
\def\ee{\end{equation}}
\def\ba{\begin{eqnarray}}
\def\ea{\end{eqnarray}}

\def\w{\wedge}
\def\d{{\rm d}}

\def\k{\kappa}
\def\r{\rho}
\def\a{\alpha}

\def\b{\beta}

\def\g{\gamma}
\def\G{\Gamma}

\def\D{\Delta}
\def\e{\epsilon}

\def\m{\mu}
\def\n{\nu}

\def\l{\lambda}

\def\s{\sigma}

\def\Ct{{\widetilde C}}
\def\Gt{{\widetilde G}}

\def\Rt{{\widetilde R}}

\def\eh{{\hat\eta}}
\def\et{{\widetilde\epsilon}}

\def\St{{\widetilde S}}

\def\IR{\relax{\rm I\kern-.18em R}}

\def\be{\begin{equation}}
\def\ee{\end{equation}}
\def\ba{\begin{eqnarray}}
\def\ea{\end{eqnarray}}

\def\tr{\,{\rm tr}\,}

\def\a{\alpha}
\def\b{\beta}
\def\g{\gamma}
\def\G{\Gamma}
\def\d{{\rm d}}

\def\e{\epsilon}

\def\m{\mu}
\def\n{\nu}
\def\r{\rho}
\def\l{\lambda}

\def\k{\kappa}
\def\s{\sigma}
\def\o{\omega}

\def\vf{\varphi}

\def\ks{{k \kern-.5em /}}
\def\es{{\e \kern-.4em /}}
\def\ds{{\partial \kern-.5em /}}
\def\Ds{{D \kern-.7em /}}

\def\R{{\cal R}}

\def\rg{\sqrt{\vert g\vert}}

\def\inv{^{\raise.15ex\hbox{${\scriptscriptstyle -}$}\kern-.05em 1}}

\def\st{{}^*}

\relax
\renewcommand{\theequation}{\thesection.\arabic{equation}}
\begin{document}
\topmargin-2.4cm
%\draft
%\preprint
%
%
%
%
\begin{titlepage}
\begin{flushright}
LPTENS-03/26\\ hep-th/0307152 \\ July 2003
\end{flushright}
\vskip 3.5cm

\begin{center}{\Large\bf Anomaly cancellation in M-theory:}\\
\vspace{2mm} {\Large\bf a critical review} \vskip 1.5cm {\bf Adel
Bilal$^{1}$ and Steffen Metzger$^{1,2}$}

\vskip.3cm
$^1$ CNRS - Laboratoire de Physique Th\'eorique,
\'Ecole Normale Sup\'erieure\\
24 rue Lhomond, 75231 Paris Cedex 05, France

\vskip.3cm
$^2$ Sektion Physik, Ludwig-Maximilians-Universit\"at\\
Munich,
Germany\\

\vskip.3cm
{\small e-mail: {\tt adel.bilal@lpt.ens.fr,
metzger@physique.ens.fr}}
\end{center}
\vskip .5cm

\begin{center}
{\bf Abstract}
\end{center}
\begin{quote}
We carefully review the basic examples of anomaly cancellation in
M-theory: the 5-brane anomalies and the anomalies on $S^1/{\bf
Z}_2$. This involves cancellation between quantum anomalies and
classical inflow from topological terms. To correctly fix all
coefficients and signs, proper attention is paid to issues of
orientation, chirality and the Euclidean continuation. Independent
of the conventions chosen, the Chern-Simons and Green-Schwarz
terms must always have the {\it same} sign. The reanalysis of the
reduction to the heterotic string on $S^1/{\bf Z}_2$ yields a
surprise: a previously neglected factor forces us to slightly
modify the Chern-Simons term, similar to what is needed for
cancelling the normal bundle anomaly of the 5-brane. This
modification leads to a local cancellation of the anomaly, while
maintaining the periodicity on $S^1$.
\end{quote}

%\noindent\rule{8cm}{.1mm}

%\noindent
%{\small e-mail: {\tt adel.bilal@lpt.ens.fr,
%metzger@physique.ens.fr}}
\end{titlepage}
\setcounter{footnote}{0}
\setcounter{page}{0}
\setlength{\baselineskip}{.7cm}
\newpage
%
% BODY
%

%%%%%%%%%%%%%%%%%%%%%%%%%%%%%%%%%%%%%%%%%%%%%%%%%%%%%%%%%%%%%%%%
\section{Introduction\label{Intro}}
%%%%%%%%%%%%%%%%%%%%%%%%%%%%%%%%%%%%%%%%%%%%%%%%%%%%%%%%%%%%%%%%

Various examples of anomaly cancellation in M-theory are based on
an interplay between quantum anomalies on even-dimensional
submanifolds and anomaly inflow from the 11-dimensional bulk
through a non-invariance of a topological integral like $S_{\rm
CS}\sim \int C\w\d C\w\d C$ or $S_{\rm GS}\sim\int G\w X_7$. For
such a cancellation to take place it is clearly crucial that the
Chern-Simons and Green-Schwarz terms have the correct signs (and
coefficients\footnote
{
Of course, the coefficient of $S_{\rm CS}$ was already
determined in \cite{CJS} from imposing supersymmetry.
}). Also, to compute the quantum anomaly correctly one
needs to know exactly the chirality of the fermion zero-modes as
well as the self-duality or anti-self-duality of antisymmetric
tensor fields on the even-dimensional submanifolds. While in
principle straightforward, tracking the various sign conventions
of the different computations existing in the literature is quite
an enterprise. One can find almost as many choices of one sign as
the other for $S_{\rm CS}$ and $S_{\rm GS}$ with rather little
correlation between both signs. While both choices are possible
(and related via $C\to -C$), the relative sign between $S_{\rm
CS}$ and $S_{\rm GS}$ must be fixed. This motivated us to
reanalyse the anomaly cancellation mechanisms in order to
determine various consistent conventions of signs and
coefficients. Moreover, we also present the Euclidean
continuations of all relevant expressions. This turns out to be
quite subtle but it is of some advantage since anomalies are
typically given in the Euclidean \cite{AGG}.

In the next section, after explaining our conventions, we write
the 11-dimensional supergravity action with several parameters
related to different choices of normalisations. We use form
language, but also write the results explicitly in components. In
addition we carefully rewrite the action in Euclidean signature.
The subtlety resides in the continuation of topological terms to
the Euclidean and it requires to fix some conventions about the
orientation of the manifolds.

In section 3, we discuss the five-brane solution,
and its bosonic zero-modes, carefully determining
the self-duality/ anti-self-duality of the 3-form, which
is most important to get the correct sign of the quantum
anomaly. Again, this is continued to Euclidean signature.

Section 4 is a quick summary of the results of ref. \cite{AGG} on
gauge and gravitational anomalies (in Euclidean signature), and
how they translate back to Minkowski space. Again, we have to be
particularly careful about prefactors and signs since we are not
simply interested in cancellation between quantum anomalies due to
different types of fields but in the cancellation between quantum
anomalies and classical inflows from $S_{\rm CS}$ and $S_{\rm
GS}$.

The anomalies of the 5-brane are discussed in section 5 and their
cancellation finally determines the consistent choices of
coefficients and signs for $S_{\rm CS}$ and $S_{\rm GS}$. Note
that there are two anomalies, the tangent bundle anomaly
\cite{DLM,W5} and the normal bundle anomaly \cite{W5,FHMM}.
Cancellation of both anomalies gives two independent checks, in
particular both correlating the signs of $S_{\rm CS}$ and $S_{\rm
GS}$.

As a further check, in section 6, we review the anomaly
cancellation in M-theory on $S^1/{\bf Z}_2$ which yields
the strongly coupled heterotic string \cite{HW}. After a
long series
of papers, a rather careful analysis \cite{BDS}, insisting
on periodicity in the circle coordinate, displayed
the basically correct mechanism in the so-called ``upstairs''
formalism, but it still contained one numerical error.
Taking this missing factor into account, anomaly cancellation
forces us to slightly modify the Chern-Simons term, much
as in the discussion of the normal bundle anomaly of the
5-brane \cite{FHMM} or of anomalies on singular $G_2$-manifolds
in \cite{BM}. Using this modified Chern-Simons term,
for the first time, we achieve anomaly cancellation
locally, i.e. on each ten-plane separately, while still
respecting periodicity on $S^1$.
In section 7 we summarise our results.

%%%%%%%%%%%%%%%%%%%%%%%%%%%%%%%%%%%%%%%%%%%%%%%%%%%%%%%%%%%%%%%%
\section{11-dimensional supergravity and the Green-Schwarz
term: Minkowskian versus Euclidean\label{action}}
\setcounter{equation}{0}
%%%%%%%%%%%%%%%%%%%%%%%%%%%%%%%%%%%%%%%%%%%%%%%%%%%%%%%%%%%%%%%%

%%%%%%%%%%%%%%%%%%%%%%%%%%%%%%%%%%%%%%%%%%%%%%%%%%%%%%%%%%%%%%%%
\subsection{Conventions in Minkowski signature}
%%%%%%%%%%%%%%%%%%%%%%%%%%%%%%%%%%%%%%%%%%%%%%%%%%%%%%%%%%%%%%%%

We begin by defining our conventions. In the Minkowskian we always
use signature $(-,+,\ldots,+)$ and label the coordinates $x^\m$,
$\m=0,\ldots D-1$. We always choose a {\it right}-handed
coordinate system such that
\be\label{d1}
\int \rg\ \d x^0\w\d x^1\w \ldots \w \d x^{D-1}
= + \int\rg\ \d^D x \ge 0  \ .
\ee
(With $x^0$ being time and for even $D$, this is a non-trivial
statement. In particular, for even $D$, if we relabelled time as
$x^0\to x^D$ then $x^1,\ldots x^D$ would be a left-handed
coordinate system!) We define the $\e$-tensor as
\be\label{d2}
\e_{01\ldots (D-1)}
=+\rg \quad \Leftrightarrow \quad \e^{01\ldots (D-1)}
=-{1\over \rg} \ .
\ee
Then \be\label{d3} \d x^{\m_1} \w
\ldots \w \d x^{\m_D} =
 - \e^{\m_1\ldots\m_D}\, \rg\ \d^D x \ .
\ee
A $p$-form $\o$ and its components are related as
\be\label{d4}
\o={1\over p!}\, \o_{\m_1\ldots\m_p}\,
\d x^{\m_1} \w \ldots \w \d x^{\m_p}
\ee
and its dual is
\be\label{d5}
\st\o = {1\over p! (D-p)!}\, \o_{\m_1\ldots\m_p}\
\e^{\m_1\ldots\m_p}_{\phantom{\m_1\ldots\m_p} \m_{p+1}\ldots\m_D}
\ \d x^{\m_{p+1}}\w\ldots\w\d x^{\m_D}\ .
\ee
We have $\st(\st\o)=(-)^{p(D-p)+1}\,\o$ and

\be\label{d6} \o\w \st\o= {1\over p!}\, \o_{\m_1\ldots\m_p}\,
\o^{\m_1\ldots\m_p}\ \rg\ \d^Dx \ . \ee Finally we note that the
components of the $(p+1)$-form $\xi=\d\o$ are given by
\be\label{d8} \xi_{\m_1\ldots\m_{p+1}}= (p+1)\,
\partial_{[\m_1}\o_{\m_2\ldots\m_{p+1}]}
\end{equation}
(where the brackets denote anti-symmetrisation with total weight
one) and that the divergence of a $p$-form is expressed as
\be\label{d7} \st\d \st\o={(-)^{D(p-1)+1}\over (p-1)!} \,
\nabla^\n \o_{\n\m_1\ldots\m_{p-1}}\, \d x^{\m_1} \w \ldots \w \d
x^{\m_{p-1}} \ . \ee

We define the curvature 2-form $R^{ab}={1\over 2} R^{ab}_{\ \ \
\n\s}\, \d x^\n\w\d x^\s$ in terms of the spin-connection
$\o^{ab}$ as $R^{ab}=\d\o^{ab} + \o^a_{\ c}\w \o^{cb}$. Here
$a,b,c = 0, \ldots D-1$ are ``flat'' indices, related to the
``curved'' ones by the $D$-bein $e^a_\m$. The torsion is $T^a=\d
e^a+ \o^a_{\ b}\w e^b$. The Riemann tensor $R^{\m\r}_{\ \ \ \n\s}$
is related to the curvature 2-form via $R^{ab}_{\ \ \ \n\s}=
e^a_\m e^b_\r R^{\m\r}_{\ \ \ \n\s}$, and the Ricci tensor is
$\R^\m_{\ \ \n} = R^{\m\r}_{\ \ \ \n\r}$ while the Ricci scalar
$\R$ is given by $\R=\R^\m_{\ \ \m}$. With this sign convention,
(space-like) spheres have $\R>0$.

%%%%%%%%%%%%%%%%%%%%%%%%%%%%%%%%%%%%%%%%%%%%%%%%%%%%%%%%%%%%%%%%
\subsection{The 11-dimensional supergravity action and
equations of motion in the Minkowskian}
%%%%%%%%%%%%%%%%%%%%%%%%%%%%%%%%%%%%%%%%%%%%%%%%%%%%%%%%%%%%%%%%

We start with the bosonic part of the 11-dimensional
Cremmer-Julia-Scherk supergravity action \cite{CJS} in the
Minkowskian \be\label{d9} S_{\rm M}^{\rm CJS}={1\over 2 \k^2} \int
\d^{11}x \rg \left( \g \R - {\a\over 48} G_{\m\n\r\s} G^{\m\n\r\s}
+{\b\over (144)^2} \e^{\m_1\ldots \m_{11}} C_{\m_1\m_2\m_3}
G_{\m_4\ldots\m_7} G_{\m_8\ldots\m_{11}} \right) \ee where
$\k\equiv \k_{11}$ is the 11-dimensional gravitational constant,
$G_{\m\n\r\s}= 4
\partial_{[\m}C_{\n\r\s]}$ and $\R$ is the Ricci scalar. With our
sign convention for the curvature we have $\g=+1$. Also, $\a$ must
be positive. A consistent choice is \be\label{d9b} \a=1 \ , \quad
\b=1 \ , \quad \g=1 \ . \ee Other authors define $\R$ with the
opposite sign convention which is accounted for by taking $\g=-1$.
Also, one can always rescale $C$ by a factor $\xi$, changing
$\a\to \xi^2 \a$ and $\b\to \xi^3\b$. While $\a$ remains positive,
such a rescaling can change the sign of $\b$. An invariant
combination is $\b^2/\a^3$ and hence \be\label{d14} {\b^2\over
\a^3}= 1 \ee (assuming, as we do, $G_{\m\n\r\s}= 4
\partial_{[\m}C_{\n\r\s]}$ or $G=\d C$. Some authors use $G=6\, \d
C$ in which case $\b^2/\a^3= 36$.) To facilitate comparison with
the literature we explicitly keep $\a,\ \b$ and $\g$ throughout
this paper, but the reader may just assume the choice (\ref{d9b})
and ignore all factors of $\a,\ \b,\ \g$.

The action (\ref{d9}) is
rewritten in form language as
\be\label{d10}
S_{\rm M}^{\rm CJS}={1\over 2 \k^2}\left(
\g \int \d^{11}x \rg\ \R
- {\a\over 2}\int G\w \st G - {\b\over 6} \int C\w G\w G \right)
\ee
where $G=\d C$. Note that the sign  in front of $\b$ has
changed with respect to (\ref{d9}) in accordance with the
minus sign in (\ref{d3}).

The equations of motion obtained either from (\ref{d9}) or
(\ref{d10}) by varying the $C$-field can be equivalently
written as
\be\label{d11}
\a\, \nabla_\m G^{\m\n\r\s} + {\b\over 2 \cdot 4!\cdot 4!}\
\e^{\n\r\s\m_1\ldots\m_8}\, G_{\m_1\ldots\m_4}
G_{\m_5\ldots\m_8} = 0
\ee
or
\be\label{d12}
\a \, \d\st G + {\b\over 2}\, G\w G = 0 \ .
\ee
The Einstein equations are
\be\label{d13}
\R_{\m\n}= {\a\over 12\g} \left( G_{\m\r\l\s}\, G_\n^{\ \r\l\s}
-{1\over 12}\, g_{\m\n}\, G_{\r\l\s\k}\, G^{\r\l\s\k} \right)
\ee
and can again be written in various ways.

The Green-Schwarz term \cite{VW,DLM}, which is of higher order in
$\k$, is written as
\be\label{d14a}
S_{\rm GS}=- \e\ {T_2\over 2\pi} \int C\w X_8
= - \e\ {T_2\over 2\pi} \int G\w X_7
\ee
where
we assumed that one can freely integrate by parts (no boundaries
or singularities), and where
\be\label{d14b}
X_8=\d X_7 = {1\over (2\pi)^3\, 4!}
\left( {1\over 8} \tr R^4 - {1\over 32} (\tr R^2)^2 \right) \ .
\ee
Here $T_2$ is shorthand for
\be\label{d14c}
T_2=\left( {2\pi^2\over \k^2}\right)^{1/3}
\ee
and will be
interpreted as the membrane tension in section 3. The numerical
parameter $\e$ will be fixed in section 5 from anomaly
cancellation. The Green-Schwarz term modifies the $C$-equation of
motion to
\begin{equation}
\a\d\st G+{\b\over2}G\wedge G+\epsilon
\left({2\k^4\over\pi}\right)^{1\over3}X_8=0.
\end{equation}

In Table \ref{table1} we give the values of $\g$, $\a$, $\b$ and
$\e$ as well as the convention relating $G$ and $\d C$ of a few
references.

\begin{table}[h]
\begin{center}
\begin{tabular}{|c|c|c|c|c|c|c|c|c|}\hline
reference &  signature & $\g$ & $\a$ & $\b$ & $\e$ &
$G$ & ${\b^2\over \a^3}$ & ${\e^3\over \b}$  \\
\hline
\hline
\cite{CJS} & $+-\ldots -$ & $-1$ & $2\k^2$ &$-2\sqrt{2}\k^3$ &   &
$\d C$    &  1 &    \\
\hline
\cite{GSW} & $-+\ldots +$ & $-1$ & $2\k^2$ &$\mp 12\sqrt{2}\k^3$&   &
6\, $\d C$& 36 &    \\
\hline
\cite{POL} & $-+\ldots +$ &  1 & 1     & 1                  &   &
$\d C $   &  1 &    \\
\hline
\cite{BDW} & $-+\ldots +$ & $-1$& 2     & $\pm 12\sqrt{2}$ &   &
6\, $\d C$& 36 &    \\
\hline
\cite{DLM} & $-+\ldots +$ &  1 & 1     & 1                  &$-1$&
$\d C$    &  1 & $-1$ \\
\hline
\cite{KAP} & $-+\ldots +$ &  1 & 4     &  $-8$
&   &
$\d C$    &  1  &    \\
\hline
\cite{HW} (2nd ref.)&$-+\ldots +$&$-1$&2&$\pm12\sqrt2$&&$\d C$&36 &\\
\hline
\cite{FHMM}&($-+\ldots +$)&   &   &$4\pi\kappa^2$&   &$\d C$   &    &    \\
\hline
\cite{FLO} &($-+\ldots +$)&    &       &$4\pi\k^2$&$-{2\pi\over T_2}$&
$\d C$    &    & $-1$ \\
\hline
\cite{BDS} & $-+\ldots +$ & $-1$& 1     & 1
& 1 &
$\d C$    &  1 &  1 \\
\hline \cite{GAUNT}&$-+\ldots +$ &  1 & 1     & 1 & 1 &
$\d C$    &  1 &  1 \\
\hline
\end{tabular}
\end{center}
\caption{\it Various conventions used in the literature. A value
for $\b$ indicated with a $\pm$ sign means that we were not able
to trace the sign convention adopted for $\e^{01\ldots 10}$ in
this reference. Note that for ref. \cite{CJS} one has to identify
$2\k^2_{\rm CJS}=\k^2$.} \label{table1}
\end{table}
\vskip 5.mm

%%%%%%%%%%%%%%%%%%%%%%%%%%%%%%%%%%%%%%%%%%%%%%%%%%%%%%%%%%%%%%%%
\subsection{Continuation to Euclidian signature}\label{Euclidean}
%%%%%%%%%%%%%%%%%%%%%%%%%%%%%%%%%%%%%%%%%%%%%%%%%%%%%%%%%%%%%%%%

While the functional integral in the Minkowskian contains
$e^{iS_{\rm M}}$, the Euclidean one contains $e^{-S_{\rm E}}$.
This implies \be\label{d15} S_{\rm M}=i\, S_{\rm E} \quad , \quad
x^0=-i\, x^0_{\rm E} \ . \ee However, for a Euclidean manifold
$M_{\rm E}$ it is natural to index the coordinates from 1 to $D$,
not from $0$ to $D-1$. One could, of course, simply write
$ix^0=x^0_{\rm E}\equiv x^D_{\rm E}$. The problem then is for even
$D=2n$ that $\d x^0_{\rm E}\w \d x^1\w\ldots\d x^{2n-1} = -\ \d
x^1\w\ldots\d x^{2n-1} \w \d x^{2n}_{\rm E}$ and if $(x^0_E,\ldots
x^{2n-1})$ was a right-handed coordinate system then $(x^1,\ldots
x^{2n}_E)$ is a left-handed one. This problem is solved by
shifting the indices of the coordinates as \be\label{d16} i\, x^0
= x^0_{\rm E} = z^1 \ , \quad x^1=z^2 \ , \quad \ldots \ , \quad
x^{D-1} = z^D \ . \ee This is equivalent to a specific choice of
an orientation on the Euclidean manifold $M_{\rm E}$. In
particular, we impose \be\label{d16a} \int \sqrt{g}\, \d z^1\w
\ldots\w\d z^D = + \int \sqrt{g}\, \d^D z \ge 0 \ . \ee Then, of
course, for any tensor we similarly shift the indices, e.g.
$C_{157}=C^{\rm E}_{268}$ and $C_{034}=i\, C^{\rm E}_{145}$. We
have $G_{\m\n\r\s}\, G^{\m\n\r\s} = G^{\rm E}_{jklm}\, G_{\rm
E}^{jklm}$ as usual, and for a $p$-form
\be\label{d17}
\o={1\over p!}\, \o_{\m_1\ldots\m_p}\,
 \d x^{\m_1}\w\ldots\w\d x^{\m_p}
={1\over p!}\, \o^{\rm E}_{j_1\ldots j_p}\,
 \d z^{j_1}\w\ldots\w\d z^{j_p} =\o^{\rm E} \ .
\ee
In particular, we have for $p=D$
\be\label{d18}
\int_{M_{\rm M}} \o = \int_{M_{\rm E}} \o^{\rm E}\ ,
\ee
which will be most important below.
 Finally, note that the Minkowski relations
(\ref{d2}) and (\ref{d3}) become
\be\label{d19}
\d z^{j_1}\w \ldots\w\d z^{j_D}=
+\, \e_{\rm E}^{j_1\ldots j_D}\, \sqrt{g}\, \d^D z
\quad {\rm with} \quad
\e_{\rm E}^{1\ldots D}={1\over \sqrt{g}} \ .
\ee
The dual of a $p$-form $\o^{\rm E}$ is defined as in (\ref{d5})
but using $\e_{\rm E}$. It then follows that
$\st(\st\o_{\rm E})=(-)^{p(D-p)}\, \o_{\rm E}$ (with an additional
minus sign with respect to the Minkowski relation) and, as in
the Minkowskian,
$\o_{\rm E}\w \st\o_{\rm E}
= {1\over p!}\, \o^{\rm E}_{j_1\ldots j_p}\,
\o_{\rm E}^{j_1\ldots j_p}\, \sqrt{g}\ \d^D z$.

We can now readily give the Euclidean action $S_{\rm E}^{\rm CJS}$
corresponding to (\ref{d10}) or (\ref{d9}) via the relations
(\ref{d15}) and (\ref{d18}): \be\label{d20} S_{\rm E}^{\rm CJS}=
{1\over 2 \k^2} \left( -\g \int \d^{11} z \, \sqrt{g}\, \R_{\rm E}
+{\a\over 2} \int G_{\rm E} \w \st G_{\rm E} +i\ {\b\over 6} \int
C_{\rm E}\w G_{\rm E} \w G_{\rm E} \right) \ee or \be\label{d21}
S_{\rm E}^{\rm CJS}= {1\over 2 \k^2} \int \d^{11} z\,\sqrt{g}\,
\left(-\g\, \R_{\rm E} +{\a\over 48} G^{\rm E}_{jklm}\, G_{\rm
E}^{jklm} + i\ {\b\over (144)^2} \e_{\rm E}^{j_1\ldots j_{11}}
C^{\rm E}_{j_1j_2j_3} G^{\rm E}_{j_4\ldots j_7} G^{\rm
E}_{j_8\ldots j_{11}} \right) \ . \ee Note that in the Euclidean
the sign in front of $\b$ is the same in (\ref{d20}) and
(\ref{d21}) in accordance with (\ref{d19}). While the kinetic
terms in the Euclidean action are real, the topological terms are
purely imaginary! Similarly, the Euclidean Green-Schwarz term is
\be\label{d26a} S^{\rm E}_{\rm GS}= i\, \e\, {T_2\over 2\pi} \int
C_{\rm E}\w X^{\rm E}_8 =i\, \e\, {T_2\over 2\pi} \int G_{\rm E}\w
X^{\rm E}_7 \ . \ee

We now give the Euclidean equations of motion as derived
from the Euclidean action. Einstein's equations do not ``see''
the topological terms and are identical in form with
(\ref{d13}), namely
\be\label{d27}
\R^{\rm E}_{ij} = {\a\over 12\g}
\left( G^{\rm E}_{iklm} G^{{\rm E}\ klm}_j
- {1\over 12}\, g^{\rm E}_{ij}\, G^{\rm E}_{klmn} G_{\rm E}^{klmn}
\right) \ .
\ee
The $C_{\rm E}$-equation of motion now is (neglecting the
contribution from the Green-Schwarz term)
\be\label{d28}
\a\, \d\st G_{\rm E} + i\, {\b\over 2}\,G_{\rm E}\w G_{\rm E} = 0
\ee
or
\be\label{d29}
\a\, \nabla_i G_{\rm E}^{ijkl} - i\, {\b\over 2\cdot 4!\cdot 4!}\,
\e_{\rm E}^{jklm_1\ldots m_8} G^{\rm E}_{m_1\ldots m_4}
G^{\rm E}_{m_5\ldots m_8} = 0 \ .
\ee

We already noticed that the topological terms in the Euclidean
action are purely imaginary. Actually it is easy to see that
an imaginary part of the Euclidean action can only come from
terms involving $\rg\, \e^{j_1\ldots j_D}$, i.e. from topological
terms. But this is exactly what is needed to obtain anomaly
cancellation from an inflow. It is well-known \cite{AGW} that
only the imaginary part of the Euclidean {\it quantum effective}
action can be anomalous. On the other hand, only the topological
terms of the classical action provide anomaly inflow and
cancellation can precisely occur if the Euclidean topological
terms are imaginary.

Suppose quite generally that under some gauge or local Lorentz
transformation the Min\-kowski classical action has a variation
\be\label{d22} \delta S_{\rm M}^{\rm cl} = \int_{M_{\rm M}^{2n}}
D_{\rm M}^{(2n)} \ee where $M_{\rm M}^{2n}$ is a lower-dimensional
manifold of dimension $2n$. As we have seen, this is equivalent to
a variation of the Euclidean action \be\label{d23} \delta S_{\rm
E}^{\rm cl} = - i \ \int_{M_{\rm E}^{2n}} D_{\rm E}^{(2n)} \ee
with \be\label{d24} D_{\rm M}^{(2n)} = D_{\rm E}^{(2n)} \equiv
D^{(2n)} \ee according to (\ref{d17}) and (\ref{d18}). As we will
recall in section 4, the anomalies of the Euclidean effective
action are of the form $\delta \G_{\rm E}\vert_{\rm anomaly} = -
i\, \int_{M_{\rm E}^{2n}} \hat I^1_{2n}$ where $\hat I^1_{2n}$ is
a $2n$-form. Thus the total variation of $\G_{\rm E}$ is
\be\label{d24a} \delta \G_{\rm E} = - i \, \int_{M_{\rm E}^{2n}}
\left( \hat I^1_{2n}+ D^{(2n)} \right) \ . \ee Continuing back to
Minkowski signature, we get \be\label{d25} \delta \G_{\rm M} =
\int_{M_{\rm M}^{2n}} \left( \hat I^1_{2n}+ D^{(2n)} \right) \ee
where now $\hat I^1_{2n}$ is rewritten in Minkowski coordinates
according to (\ref{d17}). In any case, the condition for anomaly
cancellation is the same in Euclidean and Minkowski signature:
\be\label{d26} \hat I^1_{2n}+ D^{(2n)} = 0 \ . \ee

However, there is a further subtlety that needs to be settled when
discussing the relation between the Minkowski and the Euclidean
form of the anomaly:  we have just seen that the anomaly
density $\hat I^1_{2n}$ in the
Euclidean equals the {\it corresponding} anomaly density $\hat
I^1_{2n}$ in the Minkowskian. This means that we have to know how
the chirality matrix $\g$ is continued from the Euclidean to the
Minkowskian and vice versa. This will be relevant for the
$2n$-dimensional submanifolds. The continuation of the
$\g$-matrices is dictated by the continuation of the coordinates
we have adopted (cf (\ref{d16})):
\be\label{d30}
i\, \g^0_{\rm M}=
\g^1_{\rm E} \ , \quad \g^1_{\rm M}= \g^2_{\rm E} \ ,
\quad \ldots
\quad \g^{2n-1}_{\rm M}= \g^{2n}_{\rm E} \ .
\ee
In accordance
with ref. \cite{AGG} we define the Minkowskian and Euclidean
chirality matrices $\g_{\rm M}$ and $\g_{\rm E}$ in $2n$
dimensions as
\be\label{d31}
\g_{\rm M}=i^{n-1} \g^0_{\rm M}\ldots \g^{2n-1}_{\rm M}
\quad , \quad
\g_{\rm E} = i^n \g^1_{\rm E} \ldots  \g^{2n}_{\rm E} \ .
\ee
Both $\g_{\rm M}$ and $\g_{\rm E}$ are hermitian.
Taking into account (\ref{d30}) this leads to
\be\label{d32}
\g_{\rm M}= -  \g_{\rm E} \ ,
\ee
i.e. what we call
positive chirality in Minkowski space is called negative chirality
in Euclidean space and vice versa. This relative minus sign is
somewhat unfortunate, but it is necessary to define self-dual
$n$-forms from a pair of positive chirality spinors, both in the
Minkowskian (with our convention for the $\e$-tensor) and in
the Euclidean (with the conventions of
\cite{AGG}).\footnote { Since we will take \cite{AGG} as the
standard reference for computing anomalies in the Euclidean, we
certainly want to use the same convention for $\g_{\rm E}$. On the
other hand, we have somewhat more freedom to choose a sign
convention for $\g_{\rm M}$. The definition (\ref{d31}) of
$\g_{\rm M}$ has the further advantage that in $D=10$, $\g_{\rm
M}= \g^0_{\rm M}\ldots \g^9_{\rm M}$ which is the usual convention
used in string theory \cite{GSW}. Our $\g_{\rm M}$ also agrees
with the definition of \cite{POL} in $D=2$, 6 and 10 (but
differs from it by a sign in $D=4$ and 8).}

Indeed, as is well-known, in $2n=4k+2$ dimensions, from a pair of
spinors of the same chirality one can always construct the
components of an $n$-form $H$ by sandwiching $n$ (different)
$\g$-matrices between the two spinors. In the Minkowskian we call
such an $n$-form $H^{\rm M}$ self-dual if \be\label{d33} H^{\rm
M}_{\m_1\ldots \m_n} = +\, {1\over n!}\, \e_{\m_1 \ldots \m_{2n}}
H_{\rm M}^{\m_{n+1} \ldots \m_{2n}} \ee (with $\e$ given by
(\ref{d2})) and it is obtained from 2 spinors $\psi_I$ ($I=1,2$)
satisfying $\g_{\rm M}\,\psi_I=+\psi_I$. In the Euclidean $H^{\rm
E}$ is called self-dual if (cf \cite{AGG}) \be\label{d34} H^{\rm
E}_{j_1\ldots j_n} = +\, {i\over n!}\, \e^{\rm E}_{j_1 \ldots
j_{2n}} H_{\rm E}^{j_{n+1} \ldots j_{2n}} \ee (with $\e^{\rm E}$
given by (\ref{d19})) and it is obtained from 2 spinors $\chi_I$
($I=1,2$) satisfying $\g_{\rm E}\,\chi_I=+\chi_I$. With these
conventions a self-dual $n$-form in Minkowski space continues to
an {\it anti}-self-dual $n$-form in Euclidean space, and vice
versa, consistent with the fact that positive chirality in
Minkowski space continues to negative chirality in Euclidean
space. The situation is summarised in Table \ref{table2} where
each of the four entries corresponds to any of the 3 others.

\begin{table}[h]
\begin{center}
\begin{tabular}{|c||c|c|}\hline
        & Minkowskian            & Euclidean    \\
\hline
\hline
spinors & positive chirality &  negative chirality   \\
\hline
$n$-form& self-dual          & anti-self-dual    \\
\hline
\end{tabular}
\caption{\it Correspondences between the (anti-) self-duality of
$n$-forms in $2n=4k+2$ dimensions and the chirality of the
corresponding pair of spinors are given, as well as their
Euclidean, resp. Minkowskian continuations.} \label{table2}
\end{center}
\end{table}
In conclusion: the anomaly of a positive chirality spinor
(or a self-dual $n$-form) in
Minkowski space is given by $\delta\Gamma_M=\int_{M_M^{2n}}\hat
I_{2n}^1$ (cf. (\ref{d25})) but with $\hat I_{2n}^1$ the relevant
expression of a negative chirality spinor (or an
anti-self-dual $n$-form) in Euclidean space.

%%%%%%%%%%%%%%%%%%%%%%%%%%%%%%%%%%%%%%%%%%%%%%%%%%%%%%%%%%%%%%%%
\section{The 5-brane and its zero-modes\label{fivebrane}}
\setcounter{equation}{0}
%%%%%%%%%%%%%%%%%%%%%%%%%%%%%%%%%%%%%%%%%%%%%%%%%%%%%%%%%%%%%%%%

To determine the exact coefficient of the anomalies on the
6-dimensional 5-brane world-volume, we need to know the chirality
of the fermion zero-modes and the sign in the self-duality
condition for the 3-form zero-mode living on the 5-brane,
respectively anti-5-brane. To do so, we quickly retrace these
computations \cite{STELLE,KAP}.

%%%%%%%%%%%%%%%%%%%%%%%%%%%%%%%%%%%%%%%%%%%%%%%%%%%%%%%%%%%%%%%%
\subsection{The 5-brane solution}
%%%%%%%%%%%%%%%%%%%%%%%%%%%%%%%%%%%%%%%%%%%%%%%%%%%%%%%%%%%%%%%%

The 5-brane and anti-5-brane are solutions of 11-dimensional
supergravity that preserve half of the 32 supersymmetries.
The metric is a warped metric preserving Poincar\'e invariance
on the $(5+1)$-dimensional world-volume (for flat 5-branes)
and the 4-form $G$ has a non-vanishing flux through any
4-sphere surrounding the world-volume. This is why the
5-branes are called ``magnetic'' sources. It will be enough
for us to exhibit the bosonic fields only.

We work in Minkowski space and split the coordinates into
longitudinal ones $x^\a,\ \a=0,\ldots 5$ and transverse ones
$x^m\equiv y^m,\ m=6, \ldots 10$. Then the metric is\footnote { Of
course, one should not confuse the longitudinal indices labelled
$\a,\b,\g,\ldots$ with the coefficients $\a,\ \b$ and $\g$
appearing in the action. } \be\label{t1} \d s^2=\D(r)^{-1/3}\,
\eta_{\a\b} \d x^\a \d x^\b + \D(r)^{2/3}\, \delta_{mn} \d y^\m \d
y^\n \ee where \be\label{t1a} \D(r)= 1+ {r_0^3\over r^3} \quad ,
\quad r=\left(\delta_{mn}y^m y^n\right)^{1/2} \quad , \quad r_0\ge
0 \ , \ee (with $\eta_{\a\b}={\rm diag}(-1,1,\ldots 1)$). This
gives a Ricci tensor (with our sign conventions as described in
section 2.1 and corresponding to $\g=+1$) \be\label{t2} \R^\a_{\
\b} =-f_1(r)\ \delta^\a_{\ \b} \ , \quad \R^m_{\ n} =-f_2(r)\
\delta^m_{\ n} - f_3(r)\ {y^m y^n\over r^2} \ , \ee with
\be\label{t3} f_1(r)={3\over 2} \D(r)^{-8/3}\ {r_0^6\over r^8}
\quad , \quad f_2(r)=-2 f_1(r) \quad , \quad f_3(r)=3 f_1(r)  \ .
\ee The Einstein equations (\ref{d13}) are solved by \be\label{t4}
G_{mnpq} = f(r)\ \et_{mnpqs}\ y^s \ , \qquad {\rm all\ other\ }
G_{\m\n\r\s}=0 \ , \ee with $\et_{mnpqs}$ completely antisymmetric
and $\et_{6\, 7\, 8\, 9\, 10}=+1$, provided \be\label{t5}
f(r)^2={9\over \a}\ {r_0^6\over r^{10}} \ . \ee (Note that $\a>0$
and we used $\g=1$. If $\g=-1$ one gets the same equations
(\ref{t4}) and (\ref{t5}).) The other equation of motion
(\ref{d11}) reduces to $\partial_m\left( \rg\ G^{mnpq}\right)=0$,
which is satisfied independent of the detailed form of the
function $f(r)$. Hence there are two solutions: \be\label{t6}
G_{mnpq}=\pm\ {3\, {\rm sgn}(\b)\over \sqrt{\a}}\ {r_0^3\over
r^5}\ \et_{mnpqs}\ y^s \ , \qquad {\rm all\ other\ }
G_{\m\n\r\s}=0 \ . \ee The solution with the upper sign ($+$) is
called a 5-brane and the one with the lower sign ($-$) an
anti-5-brane. If one simply redefines $C\to -C$ and hence $G\to
-G$ one also has $\b\to -\b$. With the factor of ${\rm sgn}(\b)$
inserted in (\ref{t6}) a 5-brane remains a 5-brane under {\it any}
rescaling $C\to \xi C$, even with negative $\xi$. This is
desirable since we will assign certain chiral zero-modes to the
5-brane and this chirality (and the corresponding anomaly) should
not be changed by a simple rescaling of $C$. The corresponding
4-form is \be\label{t7} G=\pm\ {\rm sgn}(\b)\ {r_0^3\over 8
\sqrt{\a}}\ \et_{mnpqs}\ {y^s\over r^5}\, \d y^m\w \d y^n\w \d
y^p\w \d y^q \ee and for any 4-sphere in the transverse space
surrounding the world-volume we have the ``magnetic charge''
\be\label{t7a} {\rm sgn}(\b)\sqrt{\a} \int_{S^4} G\ =\ \pm\ 3
r_0^3 {\rm vol}(S^4) \ =\ \pm\ 8\pi^2 r_0^3 \ . \ee Hence, for the
5-brane the flux of ${\rm sgn}(\b) G$ is positive and for the
anti-5-brane it is negative.

The parameter $r_0$ sets the scale for the (anti-) 5-brane
solution. One can compute the energy per 5-volume of the brane,
i.e. the 5-brane tension $T_5$. From (\ref{d9}) we see that it
must be ${1\over \k^2}$ times a function of $r_0$ and hence, on
dimensional grounds, $T_5\sim {r_0^3\over \k^2}$. Including the
exact numerical coefficient \cite{STELLE} gives \be\label{t8}
T_5={8\pi^2 r_0^3\over 2\k^2} \ .\ \ee Thus, the ``magnetic
charge'' of the (anti-) 5-brane can be expressed as ${\rm
sgn}(\b)\sqrt{\a} \int_{S^4} G\ =\ \pm 2\k^2 T_5$. The Dirac
quantisation condition between membranes and 5-branes then relates
the membrane tension $T_2$ and the 5-brane tension $T_5$ as
\cite{ALW1} \be\label{t9} T_2\, T_5 ={2\pi\over 2\k^2}\ \ee so
that $8\pi^2 r_0^3={2\pi\over T_2}$ and (\ref{t7a}) can be
rewritten as
\begin{equation}\label{t10}
{\rm sgn}(\b)\sqrt{\a}\int_{S^4} G\ =\ \pm\ {2\pi\over T_2}.
\end{equation}
This is equivalent to the modified Bianchi identity
\begin{equation}\label{t11}
{\rm sgn}(\b)\sqrt{\a}\ \d G = \ \pm\ {2\pi\over T_2}\
\delta^{(5)}_{W_6} \qquad \pmatrix{ + \mbox{\ for\ 5-brane}\cr -
\mbox{ \ for\ anti-5-brane}\cr}
\end{equation}
where
$\delta^{(5)}_{W_6}$ is a 5-form Dirac distribution with support
on the world-volume $W_6$ such that
%\be\label{t12}
$\int_{M_{11}} \o_{(6)}\w \delta^{(5)}_{W_6}
= \int_{W_6} \o_{(6)} \ .\
$%\ee
Finally, there is one more relation between the tensions,
namely \cite{ALW1}
\be\label{t13}
T_2^2= 2\pi T_5 \ , \
\ee
which together with (\ref{t9}) allows us to express
$T_2$ and $T_5$
in terms of $\k$ alone, in particular
\be\label{t14}
T_2=\left( {2\pi^2\over \k^2}\right)^{1/3}
\ee
as already anticipated in (\ref{d14c}).

To summarise, the 5-brane and anti-5-brane solutions both have a
metric given by (\ref{t1}). The 4-form $G$ is given by (\ref{t7})
and satisfies the Bianchi identity (\ref{t11}) with $T_2$ given
in terms of $\k$ by (\ref{t14}). The upper sign always corresponds
to 5-branes and the lower sign to anti-5-branes.

%%%%%%%%%%%%%%%%%%%%%%%%%%%%%%%%%%%%%%%%%%%%%%%%%%%%%%%%%%%%%%%%
\subsection{Zero-modes of the 5-brane}
%%%%%%%%%%%%%%%%%%%%%%%%%%%%%%%%%%%%%%%%%%%%%%%%%%%%%%%%%%%%%%%%

We will consider the zero-modes of the bosonic equations of motion
in the background of the 5-brane solution. The anti-5-brane
background can be treated similarly (flipping signs in appropriate
places).

Apart from fluctuations describing the position of the 5-brane,
there are zero-modes of the $C$-field. A zero-mode is a
square-integrable fluctuation $\delta G=\d \delta C$ around the
5-brane solution $G_0$ (given by (\ref{t6}) or (\ref{t7}) with the
upper sign) such that $G=G_0+\delta G$ still is a solution of
(\ref{d11}) or (\ref{d12}). Of course, $G$ must also solve the
Einstein equations to first order in $\delta G$. This will be the
case with the same metric if the r.h.s. of (\ref{d13}) has no term
linear in $\delta G$.

The linearisation of eq. (\ref{d11}) around the 5-brane solution
(\ref{t6}) is \be\label{t15} \nabla_\m \delta G^{\m\n\r\s} +
{\b\over 4! \, 4!\, \a}\, {3\, {\rm sgn}(\b\,) r_0^3\over
\sqrt{\a}\, r^5}\ \e^{\n\r\s\m_1\ldots \m_4 mnpq}\ \et_{mnpqs}\,
y^s \, \delta G_{\m_1\ldots\m_4} =0 \ . \ee Since there are only 5
transverse directions, the second term is non-vanishing only if
exactly one of the indices $\n\r\s\m_1\ldots \m_4$ is transverse.
It is not too difficult to see that the only solutions are such
that all components of $\delta G$ but $\delta G_{m\a\b\g}$ vanish.
This also ensures that $\delta G$ cannot contribute linearly to
the Einstein equations. We take the ansatz \cite{KAP}
\be\label{t16} \delta G_{m\a\b\g} =\D(r)^{-1-\zeta}\, r^{-5}\,
y^m\, H_{\a\b\g} \ , \quad {\rm with}\ \partial_n H_{\a\b\g}=0 \ ,
\ee and use $\rg=\D(r)^{2/3}$, $g^{mn}=\D(r)^{-2/3}\,
\delta^{mn}$, $g^{\a\b}=\D(r)^{1/3}\, \eta^{\a\b}$, as well as the
convention that indices of $H_{\a\b\g}$ are raised with
$\eta^{\a\b}$  and those of $\delta G_{m\a\b\g}$ with $g^{mn}$ and
$g^{\a\b}$. This means that $\delta G^{m\a\b\g} =
\D(r)^{-2/3-\zeta}\, r^{-5}\, y^m\, H^{\a\b\g}$. We further need
\be\label{t17} \e^{\a\b\g t\delta\e\vf mnpq}\, \et_{mnpqs} = -
{4!\over \rg}\, \delta^t_s\ \et^{\a\b\g\delta\e\vf} \ , \ee with
$\et^{\a\b\g\delta\e\vf}$ completely antisymmetric and
$\et^{012345}=-1$, i.e. $\et$ is exactly the $\e$-tensor (as
defined in (\ref{d2})) for the $(5+1)$-dimensional world-volume
with metric $\eta_{\a\b}$. Then, for $(\n,\r,\s)=(\a,\b,\g)$, eq.
(\ref{t15}) becomes\footnote{For $(\n,\r,\s)=(m,\b,\g)$ eq.
(\ref{t15}) gives $\partial_{\a}H^{\a\b\g}=0$, so that
$H_{\a\b\g}=3\ \partial_{[\a}B_{\b\g]}$, as expected.}
\begin{equation}\label{t18}
\partial_m\left( \D(r)^{-\zeta}\, r^{-5}\, y^m\right) H^{\a\b\g}
-{\vert \b\vert\over \a^{3/2} } \, {r_0^3\over 2}\,
\et^{\a\b\g\delta\e\vf}
\D(r)^{-1-\zeta}\, r^{-8}\, H_{\delta\e\vf} = 0 \ .
\ee
Equation (\ref{d14}) and $\a>0$ imply
${\vert \b\vert\over \a^{3/2} }=+1$. Since
$\ \partial_m\left( \D(r)^{-\zeta}\, r^{-5}\, y^m\right)$
$= + 3\, \zeta\, \D(r)^{-\zeta-1}\, r_0^3\, r^{-8}$ we finally get
\be\label{t19}
\zeta\ H^{\a\b\g} = {1\over 6} \  \et^{\a\b\g\delta\e\vf}\
H_{\delta\e\vf} \ .
\ee
Consistency of this equation requires either $\zeta=+1$ in which
case $H$ is self-dual (cf (\ref{d33})) or $\zeta=-1$ in which case
$H$ is anti-self-dual.

As mentioned above, the zero-modes must be square-integrable:
\be\label{t20} \infty > \int \d^{11}x \rg\, \delta
G_{m\a\b\g}\,\delta G^{m\a\b\g} ={8\pi^2\over 3} \int_0^\infty \d
r\, r^{-4}\D(r)^{-1-2\zeta} \ \int_{W_6} \d^6 x\, H_{\a\b\g}\,
H^{\a\b\g} \ . \ee The $r$-integral converges if and only if
$\zeta>0$. Thus square-integrability selects $\zeta=+1$ and,
hence, $H=\d B$ is a self-dual 3-form on the world-volume.

To summarise, in Minkowski signature, on a 5-brane, there is a
self-dual 3-form $H$ (which continues to an anti-self-dual
Euclidean 3-form $H_{\rm E}$), while on an anti-5-brane the 3-form
$H$ is anti-self-dual (and continues to a self-dual Euclidean
3-form $H_{\rm E}$). To complete the 6-dimensional
supermultiplets, we know that  the self-dual 3-form is accompanied
by two spinors of positive chirality, and the anti-self-dual
3-form by two spinors of negative chirality.

%%%%%%%%%%%%%%%%%%%%%%%%%%%%%%%%%%%%%%%%%%%%%%%%%%%%%%%%%%%%%%%%
\subsection{The Euclidean 5-brane and its zero-modes}
%%%%%%%%%%%%%%%%%%%%%%%%%%%%%%%%%%%%%%%%%%%%%%%%%%%%%%%%%%%%%%%%

Instead of determining the zero-modes of the Minkowskian
5-brane and continuing to the Euclidean in the end, one
can work in the Euclidean from the beginning. We will now
sketch this Euclidean computation and check that the
zero-modes of $G_{\rm E}$ are indeed given by an anti-self-dual
(in the Euclidean sense) 3-form on the Euclidean world-volume.

The Euclidean Einstein equations (\ref{d27}) are formally the same
as the Minkowski ones. On the other hand, eqs. (\ref{d28}) and
(\ref{d29}) are truly complex. Nevertheless, for the 5-brane
solution the imaginary part $\sim G_{\rm E}\w G_{\rm E}$ vanishes.
It follows that the Euclidean 5-brane solution is given, in
complete analogy with the Minkowski case, by
\def\zt{\tilde z}
\def\eh{\hat\e}
\ba\label{t22}
\d s_{\rm E}^2&=&\D(r)^{-1/3}\, \delta_{ij} \d z^i \d z^j
+ \D(r)^{2/3}\, \delta_{mn} \d \zt^m \d \zt^n \\
\label{t23}
G^{\rm E}_{mnpq}&=& {3\, {\rm sgn}(\b)\over \sqrt{\a}}\
{r_0^3\over r^5}\ \eh_{mnpqs}\ \zt^s \ ,
\ea
where now $f,g,h,i,j,k=1, \ldots 6$ label\footnote
{
At this point we ran out of distinct, consistent choices of
letters for indices. We hope the reader will forgive us.
}
the longitudinal
coordinates ($z$) and $m,n,p,q,s=7,\ldots 11$ the transverse
ones ($\zt$). The function $\D(r)$ is defined as before
in terms of $r=(\delta_{mn}\zt^m\zt^n)^{1/2}$. Also,
$\eh_{mnpqs}$ is completely antisymmetric with
$\eh_{7\, 8\, 9\, 10\, 11}=+1$. $G^{\rm E}$ satisfies
the same Bianchi identity as in the Minkowskian, namely
\be\label{t24}
{\rm sgn}(\b)\sqrt{\a}\ \d G^{\rm E}
= \ {2\pi\over T_2}\ \delta^{(5)}_{W_6} \ .
\ee
Of course, we obtain the Euclidean anti-5-brane by adding
an extra minus sign on the r.h.s. of (\ref{t23}) and (\ref{t24}).

Now we consider zero-modes around this Euclidean 5-brane
background (\ref{t23}). Inserting the ansatz
$\delta G^{\rm E}_{mijk}
= \D(r)^{-1-\zeta} r^{-5} \zt^m H^{\rm E}_{ijk}$ (all
other components vanish) into the linearised eq. (\ref{d29}) we
get
\be\label{t25}
\nabla_m \delta G_{\rm E}^{mijk}
- i\ {3\b {\rm sgn}(\b\,)r_0^3 \over 4! \, 4!\, \a^{3/2}r^5 }
\, 4\, \e_{\rm E}^{ijktfghmnpq}\ \eh_{mnpqs}\, \zt^s \,
\delta G_{tfgh} =0 \ .
\ee
Now eq. (\ref{t17}) is replaced by
\be\label{t26}
\e_{\rm E}^{ijktfghmnpq}\ \eh_{mnpqs}
= - {4!\over \sqrt{g}}\, \delta^t_s\ \et_{\rm E}^{ijkfgh} \ ,
\ee
where $\et_{\rm E}^{ijkfgh}$ is the
Euclidean $\e$-tensor on the 6-dimensional Euclidean world-volume
with metric $\delta_{ij}$. Then the same steps as before lead to
\be\label{t27}
\zeta\ H_{\rm E}^{ijk}
= - {i\over 6} \  \et_{\rm E}^{ijkfgh}\ H^{\rm E}_{fgh} \ .
\ee
Consistency again requires
$\zeta=\pm 1$, and square-integrability selects $\zeta=+1$. We see
that the Euclidean 3-form $H_{\rm E}$ is anti-self dual according
to our definition (\ref{d34}), in agreement with the previous
result.

%%%%%%%%%%%%%%%%%%%%%%%%%%%%%%%%%%%%%%%%%%%%%%%%%%%%%%%%%%%%%%%%
\section{Gauge and gravitational anomalies of chiral fields
\label{anomaly}}
\setcounter{equation}{0}
%%%%%%%%%%%%%%%%%%%%%%%%%%%%%%%%%%%%%%%%%%%%%%%%%%%%%%%%%%%%%%%%

This section is a summary of the results of \cite{AGG} where
the anomalies for various chiral fields where related to index
theory. Again, the emphasis is on conventions, signs and
prefactors, since in the sequel we want to add these anomalies
to the inflow from the classical action.

%%%%%%%%%%%%%%%%%%%%%%%%%%%%%%%%%%%%%%%%%%%%%%%%%%%%%%%%%%%%%%%%
\subsection{Conventions}
%%%%%%%%%%%%%%%%%%%%%%%%%%%%%%%%%%%%%%%%%%%%%%%%%%%%%%%%%%%%%%%%

This entire section is in Euclidean space of dimension $2n$,
and hence we will mostly drop the subscript ${\rm E}$. Only
at the very end we will discuss the continuation to Minkowski
space.

Of course, we use the same conventions as \cite{AGG} (except that
we label space indices $i,j,\ldots$ instead of $\m,\n,\ldots$
which we reserved for Minkowski space). The chirality matrix
$\g\equiv \g_{\rm E}$ was defined in (\ref{d31}) and the
self-duality
condition for an $n$-form in (\ref{d34}). The language of
differential forms is used throughout. For gauge theory, the gauge
fields, field strength and gauge variation are given by
\ba\label{q1}
 \nonumber A&=& A_i\,\d z^i\ ,
\quad A_i=A_i^\a\, \l^\a\ ,
\quad (\l^\a)^\dag = - \l^\a \ , \\
F&=& \d A + A^2 \ , \quad \delta_v A= {\rm D} v= \d v+[A,v] \ .
\ea
Thus $F$ is anti-hermitian and differs by an $i$ from a
hermitian field strength used by certain authors.\footnote
{
For $U(1)$-gauge theories, the usual definition of the
covariant derivative is $\partial_j+iq{\cal A}_j$, with $q$
being the charge, and
hence $A\simeq i q {\cal A}$ and $F\simeq iq{\cal F}$
where ${\cal F}=\d {\cal A}$.
}
For gravity, one
considers the spin connection $\o^a_{\ b}$ as an $SO(2n)$-matrix
valued 1-form. Similarly, the parameters $\e^a_{\ b}$ of local
Lorentz transformations (with $\e^{ab}=-\e^{ba}$) are considered
as an $SO(2n)$-matrix. Then
\be\label{q2}
R=\d\o+\o^2\ , \quad
\delta_\e e^a=-\e^a_{\ b}e^b \ , \quad
\delta_\e \o={\rm D}\e=\d\e + [\o,\e] \ .
\ee
For spin-${1\over 2}$ fermions the relevant Dirac
operator is  ($E_a^j$ is the inverse $2n$-bein)
\be\label{q3}
\Ds=E_a^j \g^a \left( \partial_j +A_j
+{1\over 4} \o_{cd,j} \g^{cd}\right) \ , \quad \g^{cd}={1\over 2}
[\g^c,\g^d] \ .
\ee

%%%%%%%%%%%%%%%%%%%%%%%%%%%%%%%%%%%%%%%%%%%%%%%%%%%%%%%%%%%%%%%%
\subsection{Index theorems}
%%%%%%%%%%%%%%%%%%%%%%%%%%%%%%%%%%%%%%%%%%%%%%%%%%%%%%%%%%%%%%%%

The simplest index is that of a positive chirality
spin-${1\over2}$ field. Define $\Ds_{1\over2}=\Ds\ {{1+\g}\over2}$
and
\begin{equation}
{\rm ind}(i\Ds_{1\over2})=\mbox{ number of zero modes of}\
i\Ds_{1\over2}- \mbox{number of zero modes of}\
(i\Ds_{1\over2})^\dagger.
\end{equation}
Then by the Atiyah-Singer index theorem
\begin{equation}
{\rm ind}(i\Ds_{1\over2})=\int_{M_{2n}}[\hat A(M_{2n})\ {\rm
ch}(F)]_{2n}
\end{equation}
where ${\rm ch}(F)=\tr\exp\left({i\over 2\pi}F\right)$ is the
Chern character and $\hat A(M_{2n})$ is the Dirac genus of the
manifold, given below. The subscript $2n$ indicates to pick only
the part which is a $2n$-form. Note that if the gauge group is
$\prod_{k} G_k$, then ${\rm ch}(F)$ is replaced by
$\prod_k {\rm ch}(F_k)$.

Another important index is that of a positive chirality
spin-${3\over2}$ field. Such a field is obtained from a positive
chirality spin-${1\over2}$ field with an extra vector index by
subtracting the spin-${1\over2}$ part. An extra vector index leads
to an additional factor for the index density
\begin{equation}
\tr \exp\left({i\over2\pi}{1\over2}R_{ab}T^{ab}\right)=\tr
\exp\left({i\over2\pi}R \right)
\end{equation}
since the vector representation is
$(T^{ab})_{cd}=\delta^a_c\delta^b_d-\delta^a_d\delta^b_c$. Hence
\begin{equation}
{\rm ind}(iD_{3\over2})=\int_{M_{2n}}\left[\hat A(M_{2n})\left(\tr
\exp\left({i\over2\pi}R \right)-1\right)\ {\rm ch}(F)\right]_{2n}.
\end{equation}

The third type of fields which lead to anomalies are self-dual or
anti-self-dual $n$-forms $H$ in $2n=4k+2$ dimensions.
Such antisymmetric tensors fields
carry no charge w.r.t. the gauge group. As discussed in section
\ref{Euclidean} a self-dual tensor can be constructed from a
pair of positive chirality spinors. Correspondingly, the index is
$\hat A(M_{2n})$ multiplied by $\tr
\exp\left({i\over2\pi}{1\over2}R_{ab}T^{ab}\right)$, where
$T^{ab}={1\over2}\gamma^{ab}$ as appropriate for the
spin-${1\over2}$ representation. Note that the trace over the
spinor representation gives a factor $2^n$ in $2n$ dimensions.
There is also an additional factor ${1\over2}$ from the chirality
projector of this second spinor and another factor ${1\over2}$
from a reality constraint ($H$ is real):
\begin{equation}
{\rm ind}(iD_A)
={1\over4}\int_{M_{2n}} \left[\hat A(M_{2n})
\tr \exp\left({i\over2\pi}{1\over4}R_{ab}\g^{ab}\right)
\right]_{2n}
={1\over4}\int_{M_{2n}}[L(M)]_{2n}.
\label{indDA}
\end{equation}
$L(M)$ is called the Hirzebruch polynomial, and the subscript on
$D_A$ stands for ``antisymmetric tensor".
(Note that, while $\hat A(M_{2n})\tr
\exp\left({i\over2\pi}{1\over4}R_{ab}\gamma^{ab}\right)$ carries
an overall factor $2^n$,
$L(M_{2n})$ has a factor $2^k$ in front of each
$2k$-form part. It is only for $k=n$ that they coincide.)

Of course, the index of a negative chirality (anti-self-dual)
field is minus that of the corresponding positive chirality
(self-dual) field. Explicitly one has
\begin{eqnarray}
{\rm ch}(F)\hskip-3.mm
&=&\hskip-3.mm\tr \exp\left({i\over2\pi}F\right)
=\tr{\bf 1}+{i\over2\pi}\tr F+\ldots
+{i^k\over k!(2\pi)^k}\tr F^k+\ldots
\label{chF}\\
\nonumber\\
\hat A(M_{2n})\hskip-3.mm
&=&\hskip-3.mm1+{1\over (4\pi)^2}{1\over12}\tr R^2
+{1\over(4\pi)^4}\left[{1\over360}\tr R^4
+{1\over288}(\tr R^2)^2\right]\nonumber\\
&&\hskip-4.mm+{1\over(4\pi)^6}\left[{1\over5670}\tr R^6
+{1\over4320}\tr R^4\tr R^2+{1\over10368}(\tr R^2)^3\right]+\ldots
\label{Aroof}\\
\nonumber\\
\hat A(M_{2n})\left(\tr {\rm e}^{{i\over2\pi}R} -1\right)
\hskip-3.mm
&=&\hskip-3.mm(2n-1)+{1\over (4\pi)^2}{2n-25\over12}\tr R^2
\nonumber\\
&&\hskip-4.mm+{1\over(4\pi)^4}\left[{2n+239\over360}\tr R^4
+{2n-49\over288}(\tr R^2)^2\right]
\nonumber\\
&&\hskip-4.mm+{1\over(4\pi)^6}\left[{2n-505\over5670}\tr R^6
+{2n+215\over4320}\tr R^4\tr R^2
%\right.\nonumber\\
%&&\left.
+{2n-73\over10368}(\tr R^2)^3\right]+\ldots\nonumber\\
&&\label{ArooftrR}\\
L(M_{2n})\hskip-3.mm
&=&\hskip-3.mm1-{1\over (2\pi)^2}{1\over6}\tr R^2
+{1\over(2\pi)^4}\left[-{7\over180}\tr R^4
+{1\over72}(\tr R^2)^2\right]
\nonumber\\
&&\hskip-4.mm+{1\over(2\pi)^6}\left[-{31\over2835}\tr R^6
+{7\over1080}\tr R^4\tr R^2-{1\over1296}(\tr R^2)^3\right]+\ldots
\label{L(M)}
\end{eqnarray}

%%%%%%%%%%%%%%%%%%%%%%%%%%%%%%%%%%%%%%%%%%%%%%%%%%%%%%%%%%%%%%%%
\subsection{The relation between anomalies in $D=2n$
and index theorems in $(2n+2)$ dimensions}
%%%%%%%%%%%%%%%%%%%%%%%%%%%%%%%%%%%%%%%%%%%%%%%%%%%%%%%%%%%%%%%%

First we need to define what we mean by the anomaly. For the time
being we suppose that the classical action is invariant (no
inflow), but that the Euclidean quantum effective action $\G_E[A]$
has an anomalous variation under the gauge transformation
(\ref{q1}) with parameter $v$ of the form
\begin{equation}
\delta_v\G_E[A]=\int\tr v\, \mathcal{G}(A).
\end{equation}
Local Lorentz anomalies are treated analogously. Note that
\begin{equation}
\delta_v\G_E[A]=\int (D_\m v)^\a{\delta\G_E[A]\over\delta
A_\m^\a}=-\int v^\a(D_\m J^\m)^\a
\end{equation}
or\footnote{Note that if $A=A^\a\l^\a$, $B=B^\b\l^\b$ and $\tr
\l^\a\l^\b=-\delta^{\a\b}$ (the $\l^\a$ are anti-hermitian) then
e.g. $\tr AB=-A^\a B^\a$ and ${\delta\over\delta A^\alpha}\int \tr
AB=-B^\a$. Hence one must define ${\delta\over\delta
A}=-\l^\a{\delta\over\delta A^\a}$ so that ${\delta\over\delta
A}\int \tr AB=B$. Another way to see this minus sign in
${\delta\over\delta A }$ is to note that $A^\a=-\tr \l^\a A$.}
\begin{equation}
\delta_v\G_E[A]=\int \tr D_\m v{\delta\G_E[A]\over\delta
A_\m}=-\int\tr  v D_\m {\delta\G_E[A]\over\delta A_\m}
\end{equation}
so that $\mathcal{G}(A)$ is identified with
$-D_\m{\delta\G_E[A]\over\delta A_\m}$ or $\mathcal{G}(A)^\a$ with
$-(D_\m J^\m)^\a$. We will refer to $\delta_v\G_E[A]$ as the
anomaly, so our anomaly is the negative integrated divergence of
the quantum current.

A most important result of \cite{AGG} is the precise relation
between the anomaly in $2n$ dimensions and index theorems in
$2n+2$ dimensions, which for the pure gauge anomaly of a positive
chirality spin-${1\over2}$ field is (eq. (3.35) of \cite{AGG}):
\begin{equation}
\delta_v\G_E^{spin {1\over2}}[A]=+{i^n\over(2\pi)^n(n+1)!}\int
Q_{2n}^1(v,A,F).
\end{equation}
The standard descent equations
$\d Q_{2n}^1=\delta_vQ_{2n+1}$ and $\d Q_{2n+1}=\tr F^{n+1}\ $
relate $Q_{2n}^1$ to the invariant polynomial $\tr F^{n+1}$.
Comparing with (\ref{chF}) we see that the pure gauge anomaly is
thus given by
$\delta_v\G_E^{spin {1\over2}}[A]=
\int I_{2n}^{1,gauge}$ with the descent equations
$\d I_{2n}^{1,gauge}=\delta_v I_{2n+1}^{gauge}$ and
$\d I_{2n+1}^{gauge}=I_{2n+2}^{gauge}$, where
$I_{2n+2}^{gauge}= -2\pi i\, [{\rm ch}(F)]_{2n+2}\,$.
This is immediately generalised to include all gauge and local
Lorentz anomalies due to all three types of chiral fields
\begin{eqnarray}
\delta\G_E[A]&=&\int I_{2n}^1\\
\d I_{2n}^1=\delta I_{2n+1}&,&
\d I_{2n+1}=I_{2n+2}\ ,
\label{-2piiindex}
\end{eqnarray}
where $I_{2n+2}$ equals $-2\pi i$ times the relevant index density
appearing in the index theorem in $2n+2$ dimensions (corrected
by a factor of $\left(-{1\over2}\right)$ in the case of the
antisymmetric tensor
field, see below). This shows that the
Euclidean anomaly is purely imaginary. It is thus convenient to
introduce $\hat I$ as $I=-i\ \hat I$ so that
\begin{eqnarray}
\delta\G_E[A]&=&-i\int \hat I_{2n}^1\label{dG=-iIhat}\\
\d\hat I_{2n}^1=\delta \hat I_{2n+1}&,&
\d\hat I_{2n+1}=\hat I_{2n+2}\ .
\label{-2piindex}
\end{eqnarray}
Explicitly we have (always for positive Euclidean chirality, respectively Euclidean self-dual forms)
\begin{eqnarray}\label{Ihathalf}
\hat I_{2n+2}^{spin{1\over2}}&=&2\pi \left[\hat A(M_{2n})\ {\rm
ch}(F)\right]_{2n+2}\\
\hat I_{2n+2}^{spin{3\over2}}&=&2\pi \left[\hat A(M_{2n})\
\left(\tr \exp\left({i\over2\pi}R\right)-1\right)\ {\rm
ch}(F)\right]_{2n+2}\label{Ihat3half}\\
\hat I_{2n+2}^{A}&=&2\pi \left[\left(-{1\over2}\right){1\over4}\
L(M_{2n})\right]_{2n+2}.\label{IhatA}
\end{eqnarray}
The last equation contains an extra factor
$\left(-{1\over2}\right)$ with respect to the index (\ref{indDA}).
The minus sign takes into account the Bose rather than Fermi
statistics, and the $1\over2$ corrects the $2^{n+1}$ to $2^n$
which is the appropriate dimension of the spinor representation on
$M_{2n}$ while the index is computed in $2n+2$ dimensions. Note
that in the cases of interest, the spin-${3\over2}$ gravitino is
not charged under the gauge group and in (\ref{Ihat3half}) the
factor of ${\rm ch}(F)$ simply equals 1.

Equations (\ref{dG=-iIhat})-(\ref{IhatA}) together with
(\ref{chF})-(\ref{L(M)}) give explicit expressions for the
anomalous variation of the Euclidean effective action. In section
2.3. we carefully studied the continuation of topological terms
like $\int \hat I_{2n}^1$ between Minkowski and Euclidean
signature. It follows from equations (\ref{d22})-(\ref{d25}) that
the anomalous variation of the Minkowskian effective action is
given directly by $\hat I_{2n}^1$:
\begin{equation}
\delta\G_M=\int_{M^M_{2n}} \hat I_{2n}^1.
\end{equation}
However, one has to remember that (with our conventions for
$\gamma_M$) the chiralities in Minkowski space and Euclidean space
are opposite. While $\hat I^1_{2n}$ corresponds to positive
chirality in the Euclidean, it corresponds to negative chirality
in Minkowski space.

To facilitate comparison with references
\cite{GSW} and \cite{FLO} we note that
\begin{equation}
I(\cite{GSW})=(2\pi)^n\hat I_{2n+2}\ \ ,\ \ I(\cite{FLO})=-\hat
I_{2n+2}.
\end{equation}
The flip of sign between $I(\cite{FLO})$ and $\hat I_{2n+2}$ is
such that $\int I^1(\cite{FLO})$ directly gives the variation
of the Minkowskian effective action for positive chirality
spinors in the Minkowskian (with our definition of $\g_{\rm M}$).

%%%%%%%%%%%%%%%%%%%%%%%%%%%%%%%%%%%%%%%%%%%%%%%%%%%%%%%%%%%%%%%%
\subsection{Tangent and normal bundle anomalies for the
5-brane zero-modes}
%%%%%%%%%%%%%%%%%%%%%%%%%%%%%%%%%%%%%%%%%%%%%%%%%%%%%%%%%%

We have seen that on the Euclidean 5-brane world volume there live
an anti-self-dual 3-form and two negative chirality spinors. While
the 3-form cannot couple to gauge fields, the spinors couple to
the ``$SO(5)$-gauge" fields of the normal bundle. This coupling
occurs via
\begin{equation}
D_i=\partial_i+{1\over4}\omega_{ab,i}\gamma^{ab}
+{1\over4}\omega_{pq,i}\gamma^{pq}
\end{equation}
inherited from the eleven-dimensional spinor. Here $a,b$ and $i$
run from 1 to 6, while $p,q = 7,\ldots 11$. Thus $\omega_{pq,i}$
behaves as an $SO(5)$-gauge field $A^\a_i$ with generators
$\l^\a\sim{1\over2}\g^{pq}$. We see that the relevant $SO(5)$
representation is the spin representation \cite{W5} and hence
($R_{pq}=d\o_{pq}+\o_{pr}\o_{rq}\equiv R_{pq}^\perp$)
\begin{eqnarray}
F=F^\a\l^\a&\leftrightarrow&{1\over4}R_{pq}^\perp\g^{pq}\\
{\rm ch}(F)&\leftrightarrow&
\tr\exp\left({i\over2\pi}{1\over4}R_{pq}^\perp\g^{pq}\right)
\equiv{\rm ch}(S(N)).
\end{eqnarray}
This trace appeared already in (\ref{indDA}), except that there
$R_{ab}$ was the curvature on the manifold (i.e. on the tangent
bundle). One has
\begin{equation}
{\rm ch}(S(N))
%=\tr\exp\left({i\over2\pi}{1\over4}R_{pq}^\perp\g^{pq}\right)
=4\left[1-{1\over(4\pi)^2}{1\over4}\tr R_\perp^2
+{1\over(4\pi)^4}\left[-{1\over24}\tr R_\perp^4
+{1\over32}(\tr R_\perp^2)^2\right]+\ldots\right].
\end{equation}
The relevant anomaly polynomial includes an extra factor
${1\over2}$ from a chirality projector (as in (\ref{indDA}))
as well as a minus sign for negative chirality. It is
($R=\tilde R + R_\perp$)
\begin{eqnarray}
\left[-{1\over2}\ \hat A(M_6)\ {\rm ch}(S(N))\right]_8
=-{2\over(4\pi)^4}&& \hskip-5.mm
\left[ {1\over360}\tr \tilde R^4
+{1\over288}(\tr \tilde R^2)^2\right.
\nonumber\\
&& \hskip-5.mm \left.-{1\over24}\tr R_\perp^4
+{1\over32}(\tr R_{\perp}^2)^2
-{1\over48}\tr \tilde R^2\tr R_\perp^2\right].
\end{eqnarray}
The part not involving $R_\perp$ is just $-2[\hat A(M_6)]_8$ and
can be interpreted as the contribution to the tangent bundle
anomaly of the two negative chirality spinors on $M_6$. Adding the
contribution of the anti-self-dual three-form, which is
$\left[-\left(-{1\over8}\right)L(M_6)\right]_8$
(evaluated using $\tilde R$) we get the anomaly
on the Euclidean 5-brane as $\delta \G_E=-i\int \hat
I_6^{1,5-brane}$ with
\begin{eqnarray}\label{I85bran}
\hat I_8^{5-brane}&=&2\pi\left[-{1\over2}\ \hat A(M_6)\
{\rm ch}(S(N))+{1\over8}L(M_6)\right]_8
\nonumber\\
&=&-X_8(\tilde R)-\hat I_8 ^{normal}
\end{eqnarray}
where $X_{8}$ is given in (\ref{d14b}) (now with
$R\rightarrow\tilde R$) and
\begin{equation}
\hat I_8 ^{normal}
={1\over(2\pi)^34!}\left[-{1\over8}\tr R_\perp^4
+{3\over32}(\tr R_\perp^2)^2
-{1\over16}\tr \tilde R^2\tr R_\perp^2\right].
\end{equation}
The part $-X_8(\tilde R)$ is called the tangent bundle anomaly and
$-\hat I_8^{normal}$ the normal bundle anomaly.

%%%%%%%%%%%%%%%%%%%%%%%%%%%%%%%%%%%%%%%%%%%%%%%%%%%%%%%%%%%%%%
\section{Anomaly cancellation for the 5-brane
\label{fivebraneanom}}
\setcounter{equation}{0}
%%%%%%%%%%%%%%%%%%%%%%%%%%%%%%%%%%%%%%%%%%%%%%%%%%%%%%%%%%%%%%

In this section we return to Minkowski space. As we have seen, the
5-brane has chiral zero-modes on its 6-dimensional world-volume
with its Minkowski anomaly given by
\begin{equation}\label{I6brane}
\delta\G_M^{1-loop}=\int_{W_6}\hat I_6^{1,5-brane}
\end{equation}
where $\hat I_6^{1,5-brane}$ is the descent of $\hat
I_8^{5-brane}$ given in (\ref{I85bran}):
$I_8^{5-brane}=-X_8(\tilde R)-\hat I_8 ^{normal}$. The tangent
bundle anomaly $-X_8(\tilde R)$ is cancelled through inflow from
the Green-Schwarz term $\sim\int G\w X_7(R)$. The latter, however,
gives $X_8(R)=X_8(\tilde R + R_\perp)$, not $X_8(\tilde R)$. The
difference, as well as the normal bundle anomaly is cancelled
through inflow from the Chern-Simons term as was shown in
\cite{W5, FHMM}.  As a result, cancellation of the total 5-brane
anomaly fixes both coefficients of the Green-Schwarz and
Chern-Simons terms. In particular, it establishes a correlation
between the two coefficients.  Moreover, as we will see,
cancellation can only occur if the sign of the anomaly due to the
five-brane zero-modes is exactly as in (\ref{I6brane}),
(\ref{I85bran}).

Let us first consider inflow from the Green-Schwarz term
(\ref{d14a}). Using the Bianchi identity (\ref{t11}) and
$\delta X_7=\d X_6^1$ we get
\begin{equation}
\delta S_{GS}
=-\e{T_2\over2\pi}\int G\wedge\d X_6^1
=\e{T_2\over2\pi}\int \d G\wedge X_6^1
={\e\over\sqrt{\a}}\ {\rm sgn}(\b)\int \delta^{(5)}_{W_6}\w X_6^1
={\e\over\sqrt{\a}}\ {\rm sgn}(\b)\int_{W_6} X_6^1\ ,
\end{equation}
where, as already noted, $X_6^1$ is $X_6^1(R)$.
This corresponds via
descent to an invariant polynomial
\begin{equation}
\hat I_8^{GS}={\e\over\sqrt{\a}}\ {\rm sgn}(\b)X_8(R).
\end{equation}

Next, inflow from the Chern-Simons term is more subtle. We review
the computation of \cite{FHMM}, again paying particular attention
to issues of signs and orientation. The two key points in
\cite{FHMM} are (i) the regularisation
\begin{equation}
\delta^{(5)}_{W_6}\rightarrow \d\rho\wedge{e_4\over2}
\end{equation}
where $\rho(r)$ rises monotonically from $-1$ at $r=0$ to 0 at
some finite distance $\tilde{r}$ from the 5-brane, and $e_4=\d
e_3$ is a certain angular form with $\int_{S^4} {e_4\over2}=1$,
and (ii) a modification of the Chern-Simons term close to the
5-brane, where $G\neq\d C$.

The regularised Bianchi identity reads
\begin{equation}
{\rm sgn}(\b)\sqrt{\a}\,\d G={2\pi\over T_2}\,\d \rho \wedge
{e_4\over2}
\end{equation}
which is solved by (requiring regularity at $r=0$ where $e_4$ is
singular)
\begin{eqnarray}
G&=&\d C + {{\rm sgn}(\b)\pi\over\sqrt{\a}\,T_2}
(2\d\rho\wedge\d B-\d\rho\wedge e_3)
\nonumber\\
&=&{{\rm sgn}(\b)\pi\over\sqrt{\a}\,T_2}\rho\, e_4
+\d\left( C-{{\rm sgn}(\b)\pi\over\sqrt{\a}\,T_2}
(\rho\, e_3+2\d\rho\w B) \right)
\nonumber\\
&\equiv& {{\rm sgn}(\b)\pi\over\sqrt{\a}\,T_2}\rho\, e_4
+\d\,\Ct \ .
\end{eqnarray}
Under a local Lorentz transformation, $\delta e_3=\d e_2^1$, and
$G$ is invariant if $\delta C=0$ and $\delta B={1\over 2}\,e_2^1$.
Note that \cite{FHMM} include the $\d\rho\wedge B$-term in $C$ and
hence get a non-trivial transformation for $C$. If we let
$\Gt=\d\Ct$ then the modified Chern-Simons term is
\begin{equation}\label{SFHMM}
\St_{CS}=-{\b\over12\k^2}\lim_{\e\to 0}
\int_{M_{11}\backslash D_{\e}W_6}\Ct\w\Gt\w\Gt
\end{equation}
where $M_{11}\backslash D_{\e}W_6$ is $M_{11}$ with a small
``tubular" region of radius $\e$ around the 5-brane world-volume
cut out. (Of course, this radius $\e$ should not be confused with
the $\e$ which is the coefficient of the Green-Schwarz term.)
Its boundary is
\begin{equation}\label{boundary}
\partial(M_{11}\backslash
D_{\e}W_6)=-S_{\e}W_6
\end{equation}
where $S_{\e}W_6$ is the 4-sphere bundle over $W_6$. Note the
minus sign that appears since the orientation of the boundary is
opposite to that of the sphere bundle.\\
Under a local Lorentz transformation $G$ and hence $\Gt$ are
invariant and
\begin{equation}\label{VarCt}
\delta \Ct =-{{\rm sgn}(\b)\pi\over\sqrt{\a}\,T_2}
\d (\rho\, e_2^1).
\end{equation}
Inserting this variation into (\ref{SFHMM}), and using $\d \Gt=0$
one picks up a boundary contribution\footnote
{
We get three minus signs, one from (\ref{SFHMM}),
(\ref{boundary}) and (\ref{VarCt}) each.
Apparently the one from (\ref{boundary}) was overlooked in
\cite{FHMM}.
}
\begin{equation}
\delta\St_{CS}=-{\b\ {\rm sgn}(\b)\pi\over12\k^2\sqrt{\a}\
T_2}\ \lim_{\e\to 0}\, \int_{S_{\e}W_6}\rho e_2^1\w\Gt\w\Gt.
\end{equation}
In $\Gt=\d C-{{\rm sgn}(\b)\pi\over\sqrt{\a}\,T_2} (\d\rho\w
e_3+\rho\, e_4-2\d \rho\w\d B)$ the terms $\sim\d\rho$ cannot
contribute to an integral over $S_{\e}W_6$. Also the contribution
of the $\d C$-terms vanishes in the limit $\e\to 0$. Hence the
only contribution comes from \cite{FHMM,BOTT}
\begin{equation}
\int_{S_\e W_6} e_2^1\w e_4\w e_4=2 \int_{W_6} p_2(NW_6)^1
\end{equation}
where $p_2(NW_6)^1$ is related  via descent to the second
Pontrjagin class $p_2(NW_6)$ of the normal bundle.
Using $\rho(0)=-1$ and (\ref{t14}) we arrive at
\begin{equation}
\delta\St_{CS}={\b\ {\rm sgn}(\b)\over6\k^2}
\left({\pi\over\sqrt{\a}\,T_2}\right)^3 \int_{W_6} p_2(NW_6)^1
={\b\ {\rm sgn}(\b)\over \a^{3/2}} \, {\pi\over 12}
\int_{W_6} p_2(NW_6)^1 \ .
\end{equation}
This corresponds to an invariant polynomial
\begin{equation}
\hat I_8^{CS}={\b\ {\rm sgn}(\b)\over\a^{3/2}}\, {\pi\over12}\,
p_2(NW_6)\ .
\end{equation}
Using
\begin{eqnarray}
{\pi\over12}\,p_2(NW_6) &=&{1\over(2\pi)^34!}
\left(-{1\over4}\tr R_{\perp}^4
+{1\over8}(\tr R_{\perp}^2)^2\right)
\nonumber\\
X_8(R)&=&X_8(\Rt)+{1\over(2\pi)^34!}
\left({1\over8}\tr R_{\perp}^4-{1\over32}(\tr R_{\perp}^2)^2
-{1\over16}\tr \Rt^2\tr R_{\perp}^2\right)
\end{eqnarray}
we find that the total inflow corresponds to
\begin{eqnarray}
\hat I_8^{GS}+\hat I_8^{CS}
&=&{\rm sgn}(\b){\e\over\sqrt{\a}}\,X_8(\Rt)
\nonumber\\
&&+{{\rm sgn}(\b)\over(2\pi)^34!\sqrt{\a}}
\left[\left({\e\over8}-{\b\over4\a}\right)\tr R_{\perp}^4
+\left({\b\over8\a}-{\e\over32}\right)(\tr R_{\perp}^2)^2
-{\e\over16}\tr \Rt^2\tr R_{\perp}^2\right].
\nonumber\\
\label{totalinflow}
\end{eqnarray}

Now it is easy to study anomaly cancellation: Invariance of the
full quantum effective action requires that the sum of
(\ref{I85bran}) and (\ref{totalinflow}) vanishes. This gives four
equations
\begin{eqnarray}
{\rm sgn}(\b){\e\over\sqrt\a}=1\ \ &,&\ \
{{\rm sgn}(\b)\over\sqrt{\a}}\left({\e\over8}-{\b\over4\a}\right)
=-{1\over8}\nonumber\\
{{\rm sgn}(\b)\over\sqrt{\a}}\left({\b\over8\a}
-{\e\over32}\right)
={3\over32}\
\ &,&\ \ -{{\rm sgn}(\b)\over\sqrt\a}{\e\over16}=-{1\over16}.
\label{cancellationcond}
\end{eqnarray}
Recall that under an arbitrary rescaling $C\rightarrow \xi C$ we
have $\sqrt\a\rightarrow |\xi|\sqrt\a$, $\ \b\rightarrow\xi^3\b$
and $\epsilon\rightarrow \xi\e$. It is satisfying to note
that all four equations (\ref{cancellationcond}) are
invariant under this rescaling.

The first equation of (\ref{cancellationcond}) ensures the
cancellation of the tangent bundle anomaly and it fixes
\begin{equation}\label{GSsign}
\e={\rm sgn}(\b)\sqrt\a.
\end{equation}
The coefficients of the Green-Schwarz and Chern-Simons terms must
have the same sign! Using the relation (\ref{d14}) this is equivalent to
$\e^3/\b=+1$.

The three other equations (\ref{cancellationcond}) ensure
cancellation of the normal bundle anomaly, and require
\begin{equation}\label{cond1}
|\b|=\a^{3/2}\ \ ,\ \ \e={\rm sgn}(\b)\sqrt\a.
\end{equation}
Of course, $|\b|=\a^{3/2}$ was already required by supersymmetry
\cite{CJS}, cf. eq. (\ref{d14}). Fortunately, the second equation
selects the same value of $\e$ as the tangent bundle anomaly, and
again $\e^3/\b=+1$.

Clearly, anomaly cancellation cannot fix the overall sign of $\b$
and $\e$ since they can be changed by redefining $C\rightarrow
-C$, but the relative sign is fixed. It is also interesting to
note that the four conditions (\ref{cancellationcond}) for anomaly
cancellation have enough structure to provide a check that we
correctly computed the sign of the one-loop anomaly: suppose we
replaced equation (\ref{I85bran}) by
\begin{equation}
\hat I^{5-brane}_8(\eta)=-\eta\,[X_8(\Rt)+\hat I_8^{normal}]\ ,
\qquad \eta=\pm 1.
\end{equation}
Then equations (\ref{cancellationcond}) would get an extra factor
$\eta=\pm 1$ on their right-hand sides, and equation
(\ref{cond1}) would
be replaced by
\begin{equation}
|\b|=\a^{3/2}\,\eta\ \ ,\qquad \e={\rm sgn}(\b)\sqrt\a\,\eta\ .
\end{equation}
This shows that $\eta=+1$, indeed.

At first sight it might seem surprising that a one-loop anomaly of
opposite sign could not be cancelled through inflow from the
Chern-Simons or Green-Schwarz terms with their signs flipped. As
we have seen, such a sign flip merely corresponds to a
redefinition of the fields and obviously cannot yield a different
inflow. Technically speaking, the factor ${\rm sgn}(\b)$ inserted
in the 5-brane Bianchi identity (\ref{t11}) ensured that the
Euclidean 5-brane always has an anti-self-dual 3-form zero-mode,
independently of the sign of $\b$. Also, the inflow involved $\b\,
{\rm sgn}(\b)=\vert\b\vert$ and $\e\, {\rm sgn}(\b) = \e\, {\rm
sgn}(\e) =\vert\e\vert$, and does not depend on the overall sign
of $\b$ and $\e$ (although it is sensitive to the relative sign,
as repeatedly emphasised). Alternatively, we could have defined
the 5-brane anomaly by $\sqrt{\a}\, \d G ={2\pi\over T_2}\,
\delta^{(5)}_{W_6}$ without the ${\rm sgn}(\b)$ inserted. Then,
for negative $\b$, the 5-brane zero-modes would have had the
opposite chirality, reversing the sign of the anomaly, but also
the inflow would involve $\b<0$ and $\e<0$ (without a factor ${\rm
sgn}(\b)$), and the final result would have been the same. Of
course, this simply corresponds to exchanging the roles of
5-branes and anti-5-branes.

In summary, cancellation of all 5-brane anomalies is very powerful
and uniquely fixes the coefficients of the Chern-Simons and
Green-Schwarz terms as in (\ref{cond1}), i.e. up to a possible
rescaling of the $C$-field. In particular, cancellation of both
the tangent  and normal bundle anomaly require that the invariant
ratio ${\e^3\over \b}$ is
\begin{equation}\label{ebratio}
{\e^3\over\b}=+1\ .
\end{equation}

%%%%%%%%%%%%%%%%%%%%%%%%%%%%%%%%%%%%%%%%%%%%%%%%%%%%%%%%%%%%%%%%%
\section{Anomaly cancellation on $S^1/{\bf Z}_2$\label{hetanom}}
\setcounter{equation}{0}
%%%%%%%%%%%%%%%%%%%%%%%%%%%%%%%%%%%%%%%%%%%%%%%%%%%%%%%%%%%%%%%%%

When M-theory is compactified on an interval, chiral fields appear
on the 10-dimensional boundary planes. Anomaly cancellation
through inflow is only possible if the spectrum is that of the
$E_8\times E_8$ heterotic string \cite{HW}.
Alternatively one can consider
M-theory on the orbifold $S^1/{\bf Z}_2$ which has two
10-dimensional fixed planes. These two formalisms are referred to
as ``downstairs" (the interval) and ``upstairs" ($S^1$ with a
${\bf Z}_2$-projection).

In this section we will show that exactly the relation
(\ref{ebratio}) between $\b$ and $\e$ ensures anomaly cancellation
in this case, as well. Furthermore, we show how the sum of
Chern-Simons and Green-Schwarz terms reduces to the
anomaly-cancelling term of the heterotic string. All this is a
non-trivial check of the coefficients $\b$ and $\e$. We will
quickly review the results of \cite{BDS} and then show that, after
correction of a numerical error in \cite{BDS} a modification of
the Chern-Simons term in the vicinity of the fixed planes is
necessary. This is similar to the above discussion of the 5-brane
normal bundle anomaly.

We start in the downstairs formalism where $S_{CS}$ and $S_{GS}$
are given by integrals over $M_{10}$ times the interval
$I=S^1/{\bf Z}_2$,
\begin{equation}
S_{CS}=-{\b\over12\k^2}\int_{M_{10}\times I}C\w G\w G\ \ ,\qquad
S_{GS}=-{\e\over(4\pi\k^2)^{1/3}}
\int_{M_{10}\times I}G\w X_7\ .
\end{equation}
Here $\k$ is the eleven-dimensional $\k$ as before. This can be
rewritten in the upstairs formalism by replacing
$\int_{I}\ldots={1\over2}\int_{S^1}\ldots$ and appropriately
identifying the fields so that the integrand is ${\bf Z}_2$-even.
Defining
\begin{equation}
\k_{\rm U}^2=2\k^2\equiv 2\k_{\rm D}^2
\end{equation}
one has
\ba
S_{CS}&=&-{\b\over12\k_{\rm U}^2}
\int_{M_{10}\times S^1}C\w G\w G\ \ ,
\nonumber\\
S_{GS}&=&-{\e\over 2^{2/3}(4\pi\k_{\rm U}^2)^{1/3}}
\int_{M_{10}\times S^1}G\w X_7\ .
\label{GSterm}
\ea
Due to the different dependence on $\k$, when written in the
upstairs formalism, the Green-Schwarz term has an extra factor of
$2^{-{2/3}}$. This will be important later on.

As one can see from the Chern-Simons term,
$C_{\bar{\m}\bar{\n}\bar{\r}}$ is ${\bf Z}_2$-odd and
$C_{\bar{\m}\bar{\n} 10}$ is ${\bf Z}_2$-even
($\bar\m,\bar\n,\ldots=0,\ldots9$). The projection on
${\bf Z}_2$-even fields then implies e.g. that
\begin{equation}
C=\tilde B \w\d x^{10}
\end{equation}
with all other components of $C$ projected out. Also, this
${\bf Z}_2$-projection only leaves half of the components of the
eleven-dimensional gravitino \cite{HW}. What remains is a
ten-dimensional gravitino of positive chirality (in the
Minkowskian), together with one negative chirality
spin-${1\over2}$ field. Of course, in the Euclidean, this
corresponds to one negative chirality spin-$3\over2$ and a
positive chirality spin-$1\over2$ fermion. The 1-loop anomaly due
to the eleven-dimensional gravitino on each 10-plane $M_{10}^A$,
$A=1,2$ is thus given by
\begin{equation}
\hat I_{12,A}^{gravitino}={1\over2}\cdot{1\over2}\left(-\hat
I_{12}^{spin{3\over2}}(R_A)+I_{12}^{spin{1\over2}}(R_A)\right)\ ,
\end{equation}
where one factor ${1\over2}$ is due to the Majorana condition and
the other factor $1\over2$ due to the ``splitting" of the anomaly
between the two fixed planes \cite{HW}. $R_A$ denotes the
curvature two-form on $M_{10}^A$ which simply is the
eleven-dimensional curvature $R$ with its components tangent to
$S^1$ suppressed. As is well known, such a polynomial has a $\tr
R^6$-piece, and one must add an $E_8$ vector multiplet in the
adjoint representation ($\Tr {\bf 1}=248$) with positive chirality
(in the Minkowskian) Majorana spinors on each 10-plane. Then on
each plane $M_{10}^A$ one has a 1-loop anomaly corresponding to
\begin{eqnarray}
\hat I_{12,A}&=&{1\over4}
\left(-\hat I_{12}^{spin{3\over2}}(R_A)
+I_{12}^{spin{1\over2}}(R_A)\right)
-{1\over2}\hat I_{12}^{spin{1\over2}}(R_A, F_A)
\nonumber\\
&=&I_{4,A}\left[X_8(R_A)+{\pi\over3}\,I_{4,A}^2\right]\ ,
\label{factoran}
\end{eqnarray}
where we used $\Tr F_A^4={1\over100}(\Tr F_A^2)^2$,
$\Tr F_A^6={1\over7200}(\Tr F_A^2)^3$ and defined\footnote
{$I_{4,A}$ is
exactly what was called $\tilde I_{4,i}$ in \cite{BDS}.
}
\begin{equation}
I_{4,A}={1\over(4\pi)^2}\left({1\over30}\Tr F_A^2
-{1\over2}\tr R_A^2\right)
\equiv{1\over(4\pi)^2}\left(\tr F_A^2-{1\over2}\tr R_A^2\right)\ .
\end{equation}
Note that in the small radius limit with $R_1=R_2=R$ one has
\ba\label{smallrad}
\left[\hat I_{12,1}+\hat I_{12,2}\right]\Big\vert_{R_1=R_2=R}
&=&\left(I_{4,1}+I_{4,2}\right)
\left[ X_8(R) +{\pi\over 3}
\left(I_{4,1}^2+I_{4,2}^2-I_{4,1}I_{4,2}\right) \right]
\nonumber\\
&\equiv&\left(I_{4,1}+I_{4,2}\right)\, \widehat X_8(R,F_1,F_2)\ ,
\ea
thanks to the algebraic identity $a^3+b^3=(a+b)(a^2+b^2-ab)$.
Here $\widehat X_8$ is the relevant 8-form that appears
in the anomaly-cancelling term of the heterotic string:
\ba\label{hetX8}
\widehat X_8(R,F_1,F_2) = {1\over (2\pi)^3 4!}
&& \hskip-6.mm \left(\ {1\over 8} \tr R^4
+ {1\over 32} (\tr R^2)^2
-{1\over 8} \tr R^2 (\tr F_1^2 + \tr F_2^2) \right.
\nonumber\\
&&\hskip-5.mm \left.+\,{1\over 4} (\tr F_1^2)^2
+{1\over 4} (\tr F_2^2)^2
- {1\over 4} \tr F_1^2 \tr F_2^2 \right)\ .
\ea

The factorised form (\ref{factoran}) of the anomaly
on each ten-plane is a necessary condition to allow
for local cancellation through inflow. Clearly, the
$I_{4,A}X_8$-term has the right form to be cancelled through inflow
from the Green-Schwarz term, provided $G$ satisfies a modified
Bianchi identity $\d G\sim\sum_{A=1,2}\delta_A\wedge I_{4,A}$,
where $\delta_A$ is a one-form Dirac distribution such that
$\int_{M_{10}\times S^1}\omega_{(10)}\wedge
\delta_A=\int_{M_{10}^A}\o_{(10)}$. This is equivalent to
prescribing a boundary value for $G$ on the boundary planes in the
down-stairs approach. Such a modified Bianchi identity is indeed necessary
to maintain supersymmetry in the coupled 11-dimensional
supergravity/10-dimensional super-Yang-Mills system \cite{HW}.
In principle, this allows us to deduce the coefficient
$-\zeta$ on the
right-hand side of the Bianchi identity in the upstairs approach.
It is given by $-(4\pi)^2 {\k_{\rm U}^2\over \l^2}$ where
$\l$ is the (unknown) Yang-Mills coupling constant.

Hence, we start with a Bianchi identity \cite{HW}
\begin{equation}\label{BI}
\d G=-\zeta \sum_{A=1,2}\delta_A\w
I_{4,A}.
\end{equation}
The variation of the Green-Schwarz term then is
\begin{equation}
\delta S_{GS}=-\ {\e\over 2^{2/3}(4\pi\k^2_{\rm U})^{1/3}}
\int_{M_{10}\times S^1}G\w\d X_6^1
=-\ {\e \ \zeta\over 2^{2/3}(4\pi\k^2_{\rm U})^{1/3}}
\sum_A\int_{M_{10}^A} I_{4,A}\w X_6^1 \ .
\end{equation}
Provided
\be\label{zetavalue}
\zeta={2^{2/3}(4\pi\k^2_{\rm U})^{1/3} \over \e} \ ,
\ee
$\delta S_{GS}$ corresponds to an invariant polynomial
\begin{equation}
\hat I_{12}^{GS}=-\sum_{A}I_{4,A}\w X_8(R_A) \ .
\end{equation}
As promised, this cancels the part of the anomaly
(\ref{factoran}) involving $X_8$. Thus, anomaly cancellation
fixes the value of $\zeta$ to be (\ref{zetavalue}), thereby
determining the value of the 10-dimensional Yang-Mills coupling
$\l$ in terms of the 11-dimensional gravitational coupling
$\k$. Although this latter aspect has drawn some attention,
one has to realise that the more interesting relation between
$\l$ and the 10-dimensional $\k_{10}$ involves the (unknown)
radius $r_0$ of the circle, similarly to the relation between the
type IIA string coupling constant and $\k$.

To study anomaly inflow from the Chern-Simons term we have to
solve the Bianchi identity for $G$ (as we did for the 5-brane).
This involves several subtleties, discussed at length in
\cite{BDS}. One important point was to respect periodicity in the
circle coordinate $x^{10}\in[-\pi r_0,\pi r_0]$ which led to the
introduction of two periodic ${\bf Z}_2$-odd ``step''
functions $\e_A(x^{10})$ such that
$\e_1(x^{10})={\rm sgn}(x^{10})-{x^{10}\over\pi r_0}$ and
$\e_2(x^{10})=\e_1(x^{10}\pm \pi r_0)$. They satisfy
\begin{equation}
{1\over2}\,\d\e_A=\delta_A-{\d x^{10}\over2\pi r_0}\ .
\end{equation}
Regularising $\e_A$ (and hence $\delta_A$) properly gives
\begin{equation}\label{eed}
\delta_A\e_B\e_C\simeq{1\over3}\, \delta_A\
\delta_{BA}\delta_{CA} \ ,
\end{equation}
where $\delta_{BA}$ and $\delta_{CA}$ denote the Kronecker symbol.
When solving the Bianchi identity (\ref{BI}) one can (locally)
trade terms ${1\over2}\e_AI_{4,A}$ for terms $-\left(\delta_A-{\d
x^{10}\over2\pi r_0}\right)\o_{3,A}$, where
\begin{equation}
\d\o_{3,A}=I_{4,A}\ \ ,\qquad \delta\omega_{3,A}=\d\o_{2,A}^1 \ ,
\end{equation}
since their difference is a total derivative ($\o_{3,A}$ is given
in terms of the Chern-Simons forms on $M_{10}^A$ and has no $\d
x^{10}$ component). This introduces an arbitrary real parameter
$b$ into the solution:
\begin{eqnarray}\label{G}
G&=&\d C-b\,{\zeta\over2}\,
\sum_A\left(\e_AI_{4,A}+ \o_{3,A}\w{\d x^{10}\over\pi r_0}\right)
+(1-b) \, \zeta\, \sum_A \delta_A \w \o_{3,A}
\nonumber\\
&=&\d\left(C-b\,{\zeta\over2}\,\sum_A\e_A\o_{3,A}\right)
+\zeta\,\sum_A\delta_A\w\o_{3,A}
\nonumber\\
&\equiv&\d\,\Ct+\zeta\sum_A\delta_A\w\o_{3,A} \ .
\end{eqnarray}
Since $G$ appears in the kinetic term $\sim\int G\w\st G$,
as well as in the energy-momentum tensor, it must be gauge
and local Lorentz invariant, $\delta G=0$. This is achieved if
\cite{BDS}
\begin{eqnarray}
\delta C&=&b\,\zeta\, \sum_{A=1,2}\o_{2,A}^1 \w
{\d x^{10}\over2\pi r_0}
+(1-b) \, \zeta \sum_A \delta_A \w \o_{2,A}^1
\nonumber\\
\Leftrightarrow\ \
\delta\Ct&=&\d\left(-b\,{\zeta\over2}\sum_A\e_A\o_{2,A}^1\right)
+\zeta\sum_A \delta_A\w\o_{2,A}^1\ .
\label{deltaC}
\end{eqnarray}

In \cite{BDS} several arguments were given in favour of one
particular value of $b$, namely $b=1$, since only then $G$ is
globally well-defined. Furthermore, the higher Fourier modes of
$C_{\bar\m\bar\n 10}$ are gauge invariant only for this value of
$b$, which is a necessary condition for a safe truncation to the
perturbative heterotic string. Last, but not least, it is only for
$b=1$ that $G$ has no terms involving $\delta_A$ which would lead
to divergent pieces in the kinetic term $\int G\w\st G$.
Nevertheless, we will keep this parameter $b$ for the time being
and show in the end that anomaly cancellation also requires $b=1$.

Note that, although $G\neq\d C$, we still have  $G=\d\Ct$
as long as
we stay away from the fixed planes. This
motivates us to introduce a modified Chern-Simons term similar to
what was done in section 5 for the 5-brane or in \cite{BM}
when discussing M-theory on singular $G_2$-manifolds. We take
\begin{equation}\label{SCStilde}
\St_{CS}=-{\b\over12\k^2_{\rm U}}\int_{M_{10}\times S^1}\Ct\w G\w
G \ ,
\end{equation}
which away from the fixed planes is just
$\sim\int \Ct\w\d\Ct\w\d\Ct$.
Then
\begin{eqnarray}\label{deltaCS}
\delta\St_{CS}&=&-{\b\over12\k_{\rm U}^2}
\int_{M_{10}\times S^1}\delta \Ct\w G\w G
\nonumber\\
&=&-{\b\over12\k^2_{\rm U}}\int_{M_{10} \times S^1}
\left[\d\left(-b\,{\zeta\over2}\sum_A\e_A\omega_{2,A}^1\right)\w
2\,\d\Ct\w\zeta\sum_C\delta_C\w\o_{3,C} \right.
\nonumber\\
&&\left. \hskip28.mm
+\zeta\sum_A\delta_A\w\o_{2,A}^1\w\d \Ct\w\d\Ct\right] \ .
\end{eqnarray}
Note that we can freely integrate by parts (we assume that
$M_{10}$ has no boundary). Furthermore, since both $\delta_A$ and
$\d C=\d \tilde B\w\d x^{10}$ always contain a $\d x^{10}$, on the
r.h.s of eq. (\ref{deltaCS}) one can replace $\d\Ct\rightarrow -
b\,{\zeta\over2}\sum_B \e_B I_{4,B}$, so that
\begin{equation}
\delta\St_{CS} =-{\b\over12\k^2_{\rm U}}\ b^2\ \left({\zeta^3\over
4}\right) \int_{M_{10} \times S^1}\sum_{A,B,C}
(2\e_A\e_B\delta_C+\delta_A\e_B\e_C)\o_{2,A}^1\w I_{4,B}\w I_{4,C}
\ .
\end{equation}
The modified Chern-Simons term contributes three terms
$\delta\,\e\,\e$. This factor of 3 was absent in \cite{BDS} where
inflow from the unmodified Chern-Simons term was computed.
Also the result of \cite{BDS} was obtained only after using
$\int_{S^1}\d x^{10}\e_A\e_B=\pi r_0(\delta_{AB}-{1\over3})$
which somewhat
obscured the local character of anomaly cancellation. Now,
however, due to the explicit $\delta_A$ one-forms, the inflow from
$\St_{CS}$ is localised on the 10-planes $M_{10}^A$. Using
(\ref{eed}) we find
\begin{equation}\label{deltaSt}
\delta \St_{CS}
=-{\b\zeta^3\over48\k^2_{\rm U}}\ b^2\
\sum_{A=1,2}\int_{M_{10}^A}\o_{2,A}^1\w I_{4,A}\w I_{4,A} \ .
\end{equation}
Upon inserting the value of $\zeta$, equation (\ref{zetavalue}),
we see that this corresponds to an invariant polynomial
\begin{equation}
\hat I_{12}^{\widetilde{CS}}=-{\b\over \e^3}\, b^2\
{\pi\over3} \sum_{A=1,2}I_{4,A}^3 \ .
\end{equation}
This cancels the remaining piece of the anomaly (\ref{factoran})
precisely if \be\label{bequ} b^2={\e^3\over \b} \ . \ee Whatever
the (real) value of $b$ is, this shows again that the coefficients
of the Green-Schwarz and Chern-Simons terms, $\e$ and $\b$ must
have the same sign. Moreover, we noted already that only $b=1$ is
consistent, so that anomaly cancellation in the present case again
confirms \be\label{epsbetaratio} {\e^3\over \b}=+1\ . \ee
Conversely, using ${\e^3\over \b}=1$ from anomaly cancellation for
the 5-brane, we can conclude that anomaly cancellation on
$S^1/{\bf Z}_2$ requires $b=1$, as argued in \cite{BDS}.\footnote
{In \cite{BDS} inflow from the unmodified Chern-Simons term was
computed. This is three times smaller than (\ref{deltaSt}). Also
the factor $2^{2/3}$ in $\zeta$ was missing, so that the overall
inflow $\delta S_{CS}$ appeared 12 times smaller. This discrepancy
remained unnoticed since the anomaly cancellation condition was
expressed as ${(4\pi)^5\k^4b^2\over12\l^6}=1$. It is only after
relating ${\lambda^2\over\k^2}$ to the coefficient of the
Green-Schwarz term that one can use ${(4\pi)^5\k^4\over\l^6}=1$
and then ${b^2\over12}=1$ clearly is in conflict with $b=1$.}

Thus we have shown that all the anomalies are cancelled {\it
locally} through inflow from the Green-Schwarz\footnote{It is
interesting to note that $\St_{GS}=-{\e\over2^{2/3}(4\pi\k^2_{\rm
U})^{1/3}}\int_{M_{10}\times S^1}\Ct\w X_8$ would have led to the
same result.} and (modified) Chern-Simons terms with exactly the
same coefficients as already selected from cancellation of the
5-brane anomalies.

Finally, it is easy to show that in the small radius limit
($r_0\to 0$) the sum $S_{GS}+\St_{CS}$ exactly reproduces the
heterotic Green-Schwarz term. In this limit $X_8(R)$ and $X_7(R)$
are independent of $x^{10}$ and have no $\d x^{10}$
components. From $C=\tilde B\w\d x^{10}$ and $\delta C$ given in
(\ref{deltaC}) we identify the correctly normalised
heterotic $B$-field as the zero
mode of $\tilde B$ times ${(4\pi)^2\over\zeta}2\pi r_0\,$:
\begin{equation}\label{BdeltaB}
B={(4\pi)^2\over\zeta}\int_{S^1}\tilde B\w\d x^{10}\ \ , \qquad
\delta B=(4\pi)^2\sum_A\o_{2,A}^1=\o_{2,YM}^1-\o_{2,L}^1
\end{equation}
where $\o_{2,YM}^1$ and $\o_{2,L}^1$ are related to $\tr F_1^2+\tr
F_2^2$ and $\tr R^2$ via descent.  Next,
using (\ref{G}) and (\ref{zetavalue}), the Green Schwarz
term (\ref{GSterm}) gives in the small radius limit
\begin{eqnarray}
S_{GS}&\rightarrow& {1\over(4\pi)^2}\int_{M_{10}}\left(\d
B-\omega_{3,YM}+\o_{3,L}\right)\w X_7\nonumber\\
&=&-{1\over(4\pi)^2}\int_{M_{10}}B\w X_8
-{1\over(4\pi)^2}\int_{M_{10}}(\o_{3,YM}-\o_{3,L})\w X_7\ .
\label{SGShet}
\end{eqnarray}
The second term is an irrelevant local counterterm: its gauge and
local Lorentz variation corresponds to a vanishing $I_{12}\,$.
Such terms can always be added and subtracted. The modified
Chern-Simons term (\ref{SCStilde}) gives (using (\ref{G})
with $b=1$, (\ref{zetavalue}), (\ref{BdeltaB}) and
integrating by parts on $M_{10}$)
\begin{equation}
\St_{CS}\rightarrow
-{\b\over \e^3}\, \sum_{A,B}\int_{M_{10}}
\left({\pi\over (4\pi)^2} B\w I_{4,A}\w I_{4,B}
-{2\pi\over3} \o_{3,A}\w I_{4,B}\w \sum_C\o_{3,C}\right)
\int_{S^1}\e_A\, \e_B\, {\d x^{10}\over 2\pi\, r_0} \ .
\end{equation}
Using again $\b=\e^3$ and the relation
\begin{equation}
\int_{S^1}\e_A\ \e_B\ {\d x^{10}\over 2\pi\, r_0}
={1\over 2} \left(\delta_{AB}-{1\over3}\right)
\end{equation}
we get
\begin{eqnarray}
\St_{CS}&\rightarrow&
-{1\over(4\pi)^2}\int_{M_{10}}B\w{\pi\over3}
\left(I_{4,1}^2+I_{4,2}^2-I_{4,1}I_{4,2}\right)
\nonumber\\
&&-{2\pi\over9}\int_{M_{10}}(\o_{3,1}+\o_{3,2})
\left(\o_{3,1}I_{4,1}+\o_{3,2}I_{4,2}
-{1\over2}\o_{3,1}I_{4,2}-{1\over2}\o_{3,2}I_{4,1}\right) \ .
\label{SCShet}
\end{eqnarray}
Again, the second term is an irrelevant counterterm. Summing
(\ref{SGShet}) and (\ref{SCShet}) we arrive at
(cf. (\ref{smallrad}))
\begin{equation}\label{Shet}
S_{GS}+\St_{CS}\rightarrow
S_{het}=-{1\over(4\pi)^2}\int_{M_{10}}B\w \hat X_8(R, F_1, F_2)+\
\mbox{local counterterms} \ ,
\end{equation}
where $\hat X_8(R, F_1,F_2)$ is the standard heterotic 8-form
given in (\ref{hetX8}). Equation (\ref{Shet}) is the correctly
normalised
heterotic anomaly-cancelling term.\footnote{In order to facilitate
comparison with \cite{GSW} we note that $\hat
X_8={1\over(2\pi)^34!}X^{GSW}_8$, and $S_{het}$ as given in
(\ref{Shet}) exactly equals minus the expression given in
\cite{GSW}. The missing minus sign in \cite{GSW} is due to a sign
error related to the subtle issues of orientation, and is
corrected e.g. when using the anomaly polynomials as given in
\cite{FLO}.}

%%%%%%%%%%%%%%%%%%%%%%%%%%%%%%%%%%%%%%%%%%%%%%%%%%%%%%%%
\section{Conclusions and discussion\label{conclusion}}
%%%%%%%%%%%%%%%%%%%%%%%%%%%%%%%%%%%%%%%%%%%%%%%%%%%%%%%%

We have studied carefully and in quite some detail the 1-loop
anomalies and their cancellation through inflow from the Chern-Simons
and Green-Schwarz terms for the ``classic'' examples of
5-branes in M-theory and M-theory on $S^1/{\bf Z}_2$, i.e.
the strongly coupled heterotic string.

To determine the exact coefficients and signs of the one-loop
anomalies, we reviewed the Euclidean results of \cite{AGG} and
discussed the subtleties related to the continuation of their
results to Minkowski space. This involved issues of orientation
and how the Euclidean and Minkowskian chiralities are related. In
the conventions we have chosen, which are the standard Minkowskian
string conventions of \cite{GSW,POL} in $D=10$, it turned out that
the Minkowski chirality equals minus the Euclidean chirality of
\cite{AGG}. As a result, we found that the anomalous variation of
the Minkowskian effective action, due to a field with {\it
negative} Minkowski chirality, is \be\label{minkvar} \delta
\Gamma_{\rm M}\Big\vert_{\rm 1-loop} =+ \int \hat I_{2n}^1 \ , \ee
where $\hat I_{2n}^1$ is related via descent to the $\hat
I_{2n+2}$ for the corresponding {\it positive} Euclidean chirality
field as given in eqs. (\ref{Ihathalf})-(\ref{IhatA}).
Furthermore, for the 5-brane, we included an {\it ab initio}
calculation of the chirality of its zero-modes, independent of the
conventions and coefficients appearing in the 11-dimensional
supergravity action.

Starting from such a general action with arbitrary coefficients
for the Chern-Simons and Green-Schwarz terms, cancellation of
both, the tangent and normal bundle anomalies of the 5-brane fixes
both signs and coefficients up to a possible rescaling of the
fields. (Of course, the Chern-Simons term was already fixed from
supersymmetry, but it is nice to see that anomaly cancellation
requires the same coefficient.) This culminated in our relation
$\e^3=+\,\b$. In particular, the Green-Schwarz and Chern-Simon
terms must have the same sign.

When looking at anomaly cancellation for the strongly coupled
heterotic string, i.e. for M-theory on $S^1/{\bf Z}_2$ in the
``upstairs'' approach, we met a surprise. Insisting on defining
all quantities properly on $S^1$ induces some subtleties
considered in \cite{BDS}, but there, a discrepancy by a
factor 12 was unnoticed. A factor of 4 is easily accounted for by
relating various coefficients between the ``upstairs'' and
``downstairs'' approaches. The missing factor 3, however,
showed that we must use inflow from a  Chern-Simons term,
modified by additional contributions induced by the non-trivial
Bianchi identity. When all this is taken into account, we obtain
cancellation of all anomalies
provided the same relation $\e^3=+\, \b$ between the
coefficients of the Green-Schwarz and (modified) Chern-Simons
terms holds. Moreover, anomaly cancellation occurs {\it locally}
on each ten-plane.

As a final check we computed the 10-dimensional anomaly cancelling
term of the heterotic string. It receives contributions from both
the Chern-Simons and the Green-Schwarz term. The standard $\hat
X_8$ of the heterotic string \cite{GSW} can only be obtained if
the coefficients satisfy $\e^3=\b$.

In Table \ref{table3}, we summarise the different constraints
between the coefficients and where they come from.

\vskip 3.mm
\begin{table}[h]
\begin{center}
\begin{tabular}{|c||c|}\hline
origin       & relation           \\
\hline
\hline
supersymmetry  &  $|\b|=\a^{3/2}$  \\
\hline
5-brane tangent bundle anomaly & $\e=\a^{1/2}\, {\rm sgn}(\b)
\ \ \Rightarrow \ \  \e^3=\a^{3/2\,} {\rm sgn}(\b)  $ \\
\hline
5-brane normal bundle anomaly &  $|\b|=\a^{3/2}\ $ and
$\ \e^3=\a^{3/2}\, {\rm sgn}(\b) \ \ \Rightarrow \ \ \e^3=\b$  \\
\hline
$S^1/{\bf Z}_2$       & $\e^3=\b$ \\
\hline\end{tabular}
\caption{\it Shown are the different constraints on the
coefficients and where they come from.} \label{table3}
\end{center}
\end{table}

Furthermore, we want to point out that the gauge and mixed
anomalies that arise from compactification of M-theory on singular
$G_2$-manifolds have a fixed relative sign. These anomalies can
only be cancelled through inflow from the Chern-Simons and
Green-Schwarz terms if both these terms have the same sign
\cite{BM}.

Taking these constraints into account, a single free parameter
remains, related to the possibility of rescaling the $C$-field.
Hence, up to such rescalings, all coefficients are fixed as
$\a=\b=\e=1$, and the low-energy effective action of M-theory
in Minkowski space is given
in terms of the 11-dimensional supergravity action
$S_{\rm CJS}$ and the Green-Schwarz term $S_{\rm GS}$ as
\ba\label{c1}
S_{\rm M-theory}&=&S_{\rm CJS} + S_{\rm GS} + \ldots
\nonumber\\
&&
\nonumber\\
S_{\rm CJS}&=&{1\over 2 \k^2}\left(
\int \d^{11}x \rg\ \R
- {1\over 2}\int G\w \st G - {1\over 6} \int C\w G\w G \right)
\nonumber\\
&&
\nonumber\\
S_{\rm GS}&=&- \ {T_2\over 2\pi} \int C\w X_8 = - \ {T_2\over
2\pi} \int G\w X_7 = -\, {1\over(4\pi\k^2)^{1/3}}  \int G\w X_7 \
, \ea where $+\ldots$ indicates further terms of order higher than
$1\over\k^2$, such as those arising from the modifications of the
Chern-Simons term discussed in sections 5 and 6.

%%%%%%%%%%%%%%%%%%%%%%%%%%%%%%%%%%%%%%%%%%%%%%%%%%%%%%%%%%%%%%%%
\vskip 2.cm
\noindent
{\bf\large Acknowledgements}
%%%%%%%%%%%%%%%%%%%%%%%%%%%%%%%%%%%%%%%%%%%%%%%%%%%%%%%%%%%%%%%%
\vskip 3.mm

\noindent
Steffen Metzger gratefully acknowledges support
by the Gottlieb Daimler- und Karl Benz-Stiftung.
We would like to thank
Ruben Minasian for helpful discussions.

\vskip 2.cm

%%%%%%%%%%%%%%%%%%%%%%%%%%%%%%%%%%%%%%%%%%%%%%%%%%%%%%%%%%%%
%\newpage

\end{document}